\documentclass[twocolumn]{aastex63}

\received{2025 March 20}
\revised{2025 June 4}
\accepted{2025 June 5}
\submitjournal{ApJ}

\shorttitle{Secular resonances \& terrestrial planets in binary stars. II.}
\shortauthors{Haghighipour \& Andrews}

\begin{document}

\title{Secular Resonances in Planet-Hosting Binary Stars. II. 
Application to Terrestrial Planet Formation}

\correspondingauthor{Nader Haghighipour}
\email{nader@psi.edu}

\author[0000-0002-5234-6375]{Nader Haghighipour}
\affiliation{Planetary Science Institute, Tucson, AZ, USA}
\affiliation{Institute for Astronomy, University of Hawaii-Manoa, Honolulu, HI, USA}
\affiliation{Institute for Advanced Planetary Astrophysics, Honolulu, HI, USA}

\author{Michael Andrews}
\affiliation{University of Hawaii-Manoa, Honolulu, HI USA}

\begin{abstract}
Continuing our study of the effects of secular resonances on the formation of terrestrial planets in moderately close 
binary stars, we present here the results of an extensive numerical simulations of the formation of these objects. 
Considering a binary with two giant planets and a protoplanetary disk around its primary star, we have simulated the 
late stage of terrestrial planet formation for different types of the secondary, and different orbital elements of the 
binary and giant planets. Results demonstrate that terrestrial planet formation can indeed proceed constructively in such 
systems; however, as predicted by the general theory, secular resonances are suppressed and do not contribute to the 
formation process. Simulations show that it is in fact the mean-motion resonances of the inner giant planet that drive 
the dynamics of the protoplanetary disk and the mass and orbital architecture of the final bodies. 
Simulations also show that in the majority of the cases, the final systems contain only one terrestrial planet with a mass 
of 0.6--1.7 Earth masses. Multiple planets appear on rare occasions in the form of Earth--Mars analogs with the smaller planet 
in an exterior orbit. When giant planets are in larger orbits, the number of these double-planet systems increases and their 
planets become more massive. Results also show that when the orbits of the giant planets carry inclinations, while secular 
resonances are still suppressed, mean-motion resonances are strongly enhanced, drastically reducing the efficacy of 
the formation process. We present the results of our simulations and discuss their implications.
\end{abstract}


\section{Introduction}

A survey of the currently known circumstellar planets in moderately close binary stars (i.e, binaries with 
separations/periastrons smaller than 40 au) indicates that these binaries host planets ranging from terrestrial class 
to gas giants\footnote{We refer the reader to $<$https://exoplanet.eu/planets\_binary/$>$ for a complete 
and up-to-date archive of these systems.}. While the origin and formation of these planets are still subjects of investigation
\citep[a number of studies have demonstrated that circumstellar planets cannot form in situ in these binaries; see]
[for a complete review]{Haghighipour10a,Thebault15}\footnote{As shown on the above website, in the majority of the moderately 
close binaries, the binary semimajor axis is smaller than 25 au and the periastron distance varies between 5 and 13 au. Such 
orbital architecture implies strong perturbation from the farther stellar component. As shown by \citet{Artymowicz94}, the 
perturbation of this star can remove planet-forming material by truncating the disk around the planet-hosting star, and may 
also hinder the growth by causing collisions to result in shattering and fragmentation \citep{Thebault15}.}, the mere existence 
of these bodies, combined with their diversity, implies that if the formation process can occur and proceed constructively, 
planets of different sizes may form in this dynamically complex environment. 
This, combined with the fact that the formation time is different for planets of different sizes, raises 
the question of how the perturbation of large bodies impacts the formation of small objects. The purpose of this two-article 
series is to answer the above question. Specifically, our goal is to investigate how the secular resonances of giant planets 
affect the formation, composition, and orbital assembly of terrestrial planets around one star of a moderately close binary. 
In the first paper \citep[][hereafter Paper I]{Haghighipour25}, we presented the general theory of secular resonances in planet-hosting 
binary systems, and demonstrated the validity of our theory by comparing its predictions with the results of numerical simulations.
In this paper, we extend our project to the study of the effects of secular resonances on the last stage of 
terrestrial planet formation.

Whether terrestrial planets can form in moderately close binaries has been investigated by several authors
\citep{Haghighipour06,Haghighipour07,Quintana07,Guedes08}. Among these studies, \citet{Haghighipour07} present the only 
investigation where the perturbation of a giant planet is also included (all other studies include only the effect of
the secondary star). These authors studied the combined effect of the secondary star and a Jupiter-mass 
planet on the late stage of terrestrial planet formation, and identified regions of the parameter space where terrestrial 
planets could form. Although the work of \citet{Haghighipour07} presents a thorough analysis of the parameters involved
(i.e., the orbital elements of the planet and binary, as well as the binary mass ratio),
it does not differentiate between the ways that the perturbation of the giant planet affects the terrestrial
planet formation process (e.g., mean-motion or secular resonances, planet--embryo scattering, embryo--embryo interaction, etc.). 
In this paper, we extend the work of \citet{Haghighipour07} to systems with two giant planets and focus on the effects of 
their secular resonances. This choice of a system with two giant planets will also allow us to 
examine the influence of the secondary star on the mechanics of the formation process  by comparing the results with the results
of the simulations of the formation of terrestrial planets in our solar system when subject to the secular resonances of Jupiter 
and Saturn.

To study the effects of secular resonances, we follow \citet{Haghighipour16}. These authors present 
a comprehensive study of the effects of secular resonances on the formation of terrestrial planets in our
solar system. Motivated by the work of \citet{Levison03}, these authors simulated the late stage of terrestrial planet 
formation for different spatial distributions of protoplanetary bodies while subject 
to the perturbation of Jupiter and Saturn. Their results demonstrated that Jupiter's secular resonance 
$(\nu_5)$ does not play a significant role in the growth of planetary embryos (as mentioned in Paper I,
the effect of this resonance is suppressed by the perturbation of Saturn) whereas the secular 
resonance of Saturn $(\nu_6)$ is the strongest resonance, causing many planetesimals and planetary embryos to 
be ejected from the protoplanetary disk and/or to collide with the forming terrestrial planets. Their simulations 
also demonstrated that, as the protoplanetary disk loses mass, the $\nu_6$ resonance moves inward until it stops beyond 
2.1 au forcing terrestrial planets to form interior to this boundary. In this paper, we simulate the late stage
of terrestrial planet formation in all systems of Paper I and follow the same approach as that of \citet{Haghighipour16}.
We will present a detailed analysis of the results and determine the contributions of secular resonances to the final mass 
and orbital assembly of planets around the planet-hosting star.

The structure of this paper is as follows. In Section 2, we present the details of our systems and the initial 
conditions of our simulations. Section 3 analyzes the results and compares them with theoretical
predictions (Paper I). We also present in this section a comparison of our results with the simulations of terrestrial planet 
formation in our solar system. Section 4 concludes this study by summarizing our findings and discussing their implications.

\section{Details of the model and initial setup}

Following Paper I, we consider a binary with a semimajor axis ranging from 20 to 50 au. In Paper I, the binary eccentricity 
was varied between 0 and 0.5. In this paper, to explore the effect of the eccentricity, we vary this quantity between 0 and 0.7.
Because we are interested in the formation of terrestrial planets, 
and because models of terrestrial planet formation have been developed primarily for the formation of these bodies 
around the Sun, we assume, with no loss of generality, that the planet-hosting star (hereafter, the primary) 
is of solar type and has the mass of the Sun. The secondary star is taken to be of M, K, G, or F type with a mass 
equal to 0.4, 0.7, 1 or 1.3 solar masses, respectively. 

We assume two giant planets orbiting the primary star. The inner planet is taken to be of Jupiter mass and the mass of 
the outer planet is set to that of Saturn. We consider the semimajor axis of the outer planet to be at
2.5, 2.94, 3.5 or 4 au. For each value of this quantity, we place the inner planet in an orbit near a 5:2 commensurability 
similar to that of Jupiter and Saturn. The latter corresponds to the inner planet's semimajor axis being at 1.36, 1.6, 1.9, or 2.17 au, 
respectively. The orbital inclinations of the two planets are varied independently between $0^\circ$ and $10^\circ$ with respect to the plane 
of the binary. Other orbital elements of the planets are set to zero. In choosing the orbital elements of the two giant planets,
care was taken to ensure that both planets would be inside the stability region around the primary star, and
safe from the secondary's perturbation. The latter is equivalent to the implicit assumption that after truncation due 
to the perturbation of the secondary star, the protoplanetary disk around the primary maintained enough material for the
formation of giant planets to proceed constructively. 

To explore the possibility of the formation of terrestrial planets interior to the inner giant planet, we assume that after 
the two giant planets have formed, the primary star has retained a disk of planetesimals and planetary embryos extending from 
0.5 to 1.5 au. Because this star is of solar type, we set up this disk similar to those frequently used in the simulations of 
the late stage of terrestrial planet formation in our solar system. We consider 120 Moon- to Mars-sized planetary embryos 
placed randomly at 5 -- 10 mutual Hill's radii, and 550 planetesimals randomly distributed throughout the disk. The masses 
of these bodies are chosen such that the disk's surface density would follow an $r^{-1.5}$ profile.

\section{Numerical Integrations and Results}

We integrated the full system consisting of the binary, giant planets, and the protoplanetary disk for 100 Myr.
Integrations included mutual gravitational interactions among all bodies (the stars, giant planets, and all planetary embryos)
except for the gravitational forces of the individual planetesimals. We did not consider the gravitational interaction of 
each planetesimal with other planetesimals or with the giant planets and planetary embryos. 

To carry out integrations, we used a special-purpose integrator designed specifically for orbital integrations in binary 
star systems \citep{Chambers02}.
This integrator has the ability to integrate systems with two massive central bodies, symplecticly, and can
resolve close encounters between gravitationally interacting bodies using a hybrid symplectic approach.
We set the time steps of integrations to 7 days, approximately 1/20 of the shortest orbital period in the system, and simulated 
growth through perfect merging. Although unrealistic \citep[see][for more details]{Haghighipour22}, 
this assumption allows to maintain focus on the effects of the underlying 
physical processes such as the perturbations of secular resonances by avoiding computational complexities due to 
fragmentation, shattering, and collisional debris.

We carried out simulations for different values of the mass of the secondary star and a large number of combinations of the semimajor axis, 
eccentricity, and inclination of the binary and the two giant planets. Figure 1 shows the snapshots of the simulations for
four coplanar systems in a circular, 40 au binary, and for different secondary stars. In each simulation, 
the primary star (not shown) is at the origin, and the secondary star is at its corresponding semimajor axis. The red 
circles in each panel denote the planetary embryos and their growth. The gray circles are the planetesimals.
In all simulations, the giant planets, shown by black circles, were initially at 1.6 and 2.94 au. 
As mentioned in the previous section, 
the disk contained 120 planetary embryos and 550 planetesimals, shown slightly above the embryos at $t=0$. The panel at
$t=0.1$ Myr shows the eccentricity spikes due to mean-motion resonances with the inner planet (see Table 4 for the exact locations
of these resonances). The numbers above the
red circles in the $t=100$ Myr panels show the final mass of each object in terms of the mass of Earth.
The secular resonances of the two giant planets are at ${a_{g_1}} = 0.31-0.38$ au and  ${a_{g_2}} = 0.65-0.7$ au.
That the secular resonances are suppressed (see Paper I for the relevant analysis) can be seen at $t=0.1$ Myr in all simulations.

Table 1 shows the results of all our coplanar simulations with the giant planets at 1.6 and 2.94 au. In this table, 
$({a_{\rm b}},{e_{\rm b}},{q_{\rm b}})$ represent the semimajor axis, eccentricity, and periastron distance of the binary, and
$({a_{g_1}} , {a_{g_2}})$ are the initial semimajor axes of the secular resonances (we refer the reader to Paper I for details
on the identification of these resonances). The columns $N$ to $T_{\rm p}$ in this table present the final results. Here,
the quantities $({m_{\rm p}} , {a_{\rm p}}, {e_{\rm p}} , {i_{\rm p}})$ correspond to the planet's mass, semimajor axis, eccentricity, 
and orbital inclination, $N$ is the number of the final planets, and $\Delta {a_{\rm p}}$ denotes the planet's displacement in 
semimajor axis with respect to the initial location of its seed embryo. A negative/positive $\Delta {a_{\rm p}}$ refers to an 
inward/outward displacement of the body, meaning that the growth started farther out/in and the body moved in/out while it was 
growing. The last column, $T_{\rm p}$, shows the time of the last major impact (i.e., embryo--embryo collision).

\begin{figure*}[ht]
\center
\vskip 12pt
\hskip -14pt
\includegraphics[scale=0.35 , angle=-90]{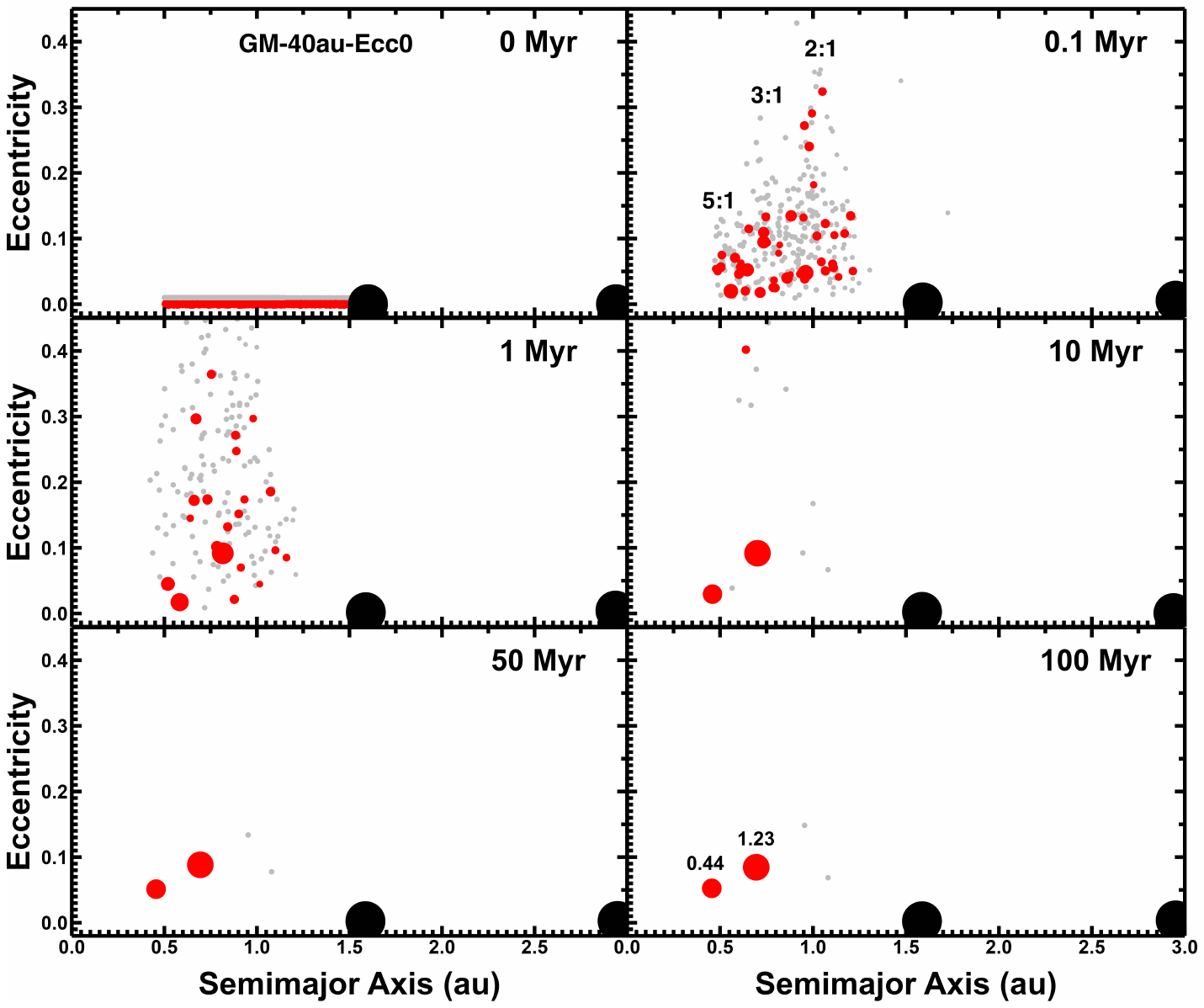}
\hskip -33pt
\includegraphics[scale=0.35 , angle=-90]{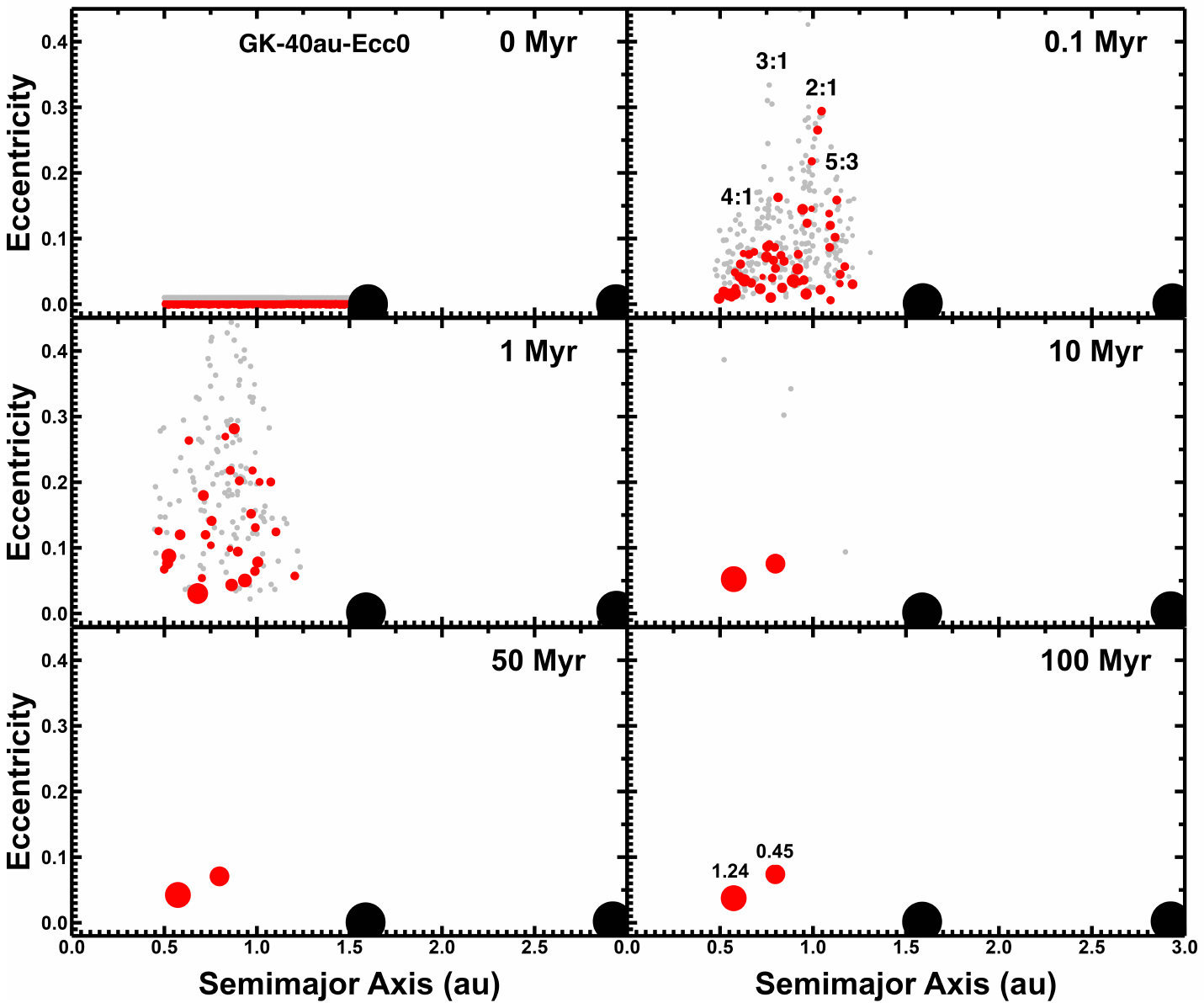}
\vskip -5pt
\hskip -14pt
\includegraphics[scale=0.35 , angle=-90]{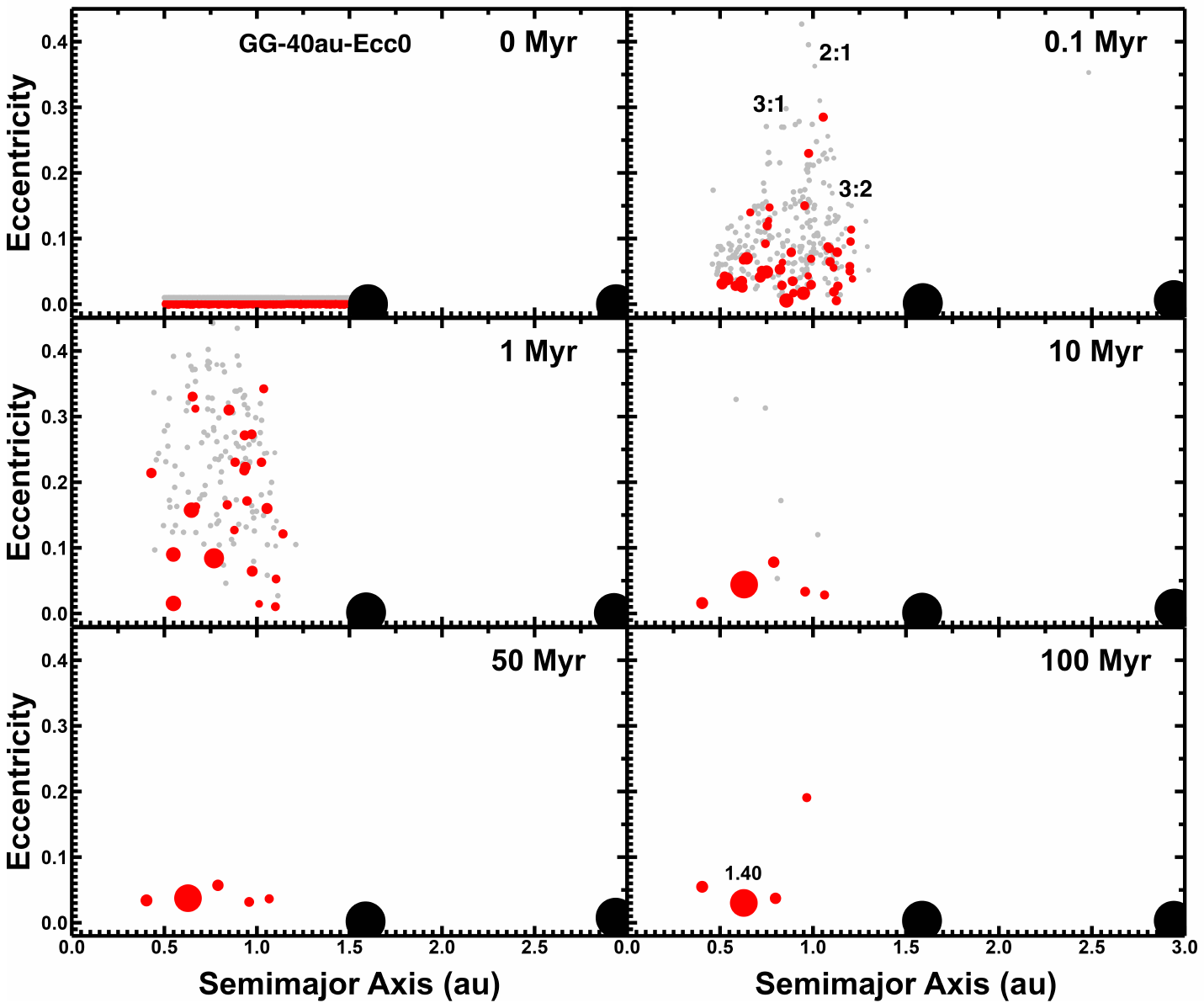}
\hskip -33pt
\includegraphics[scale=0.35 , angle=-90]{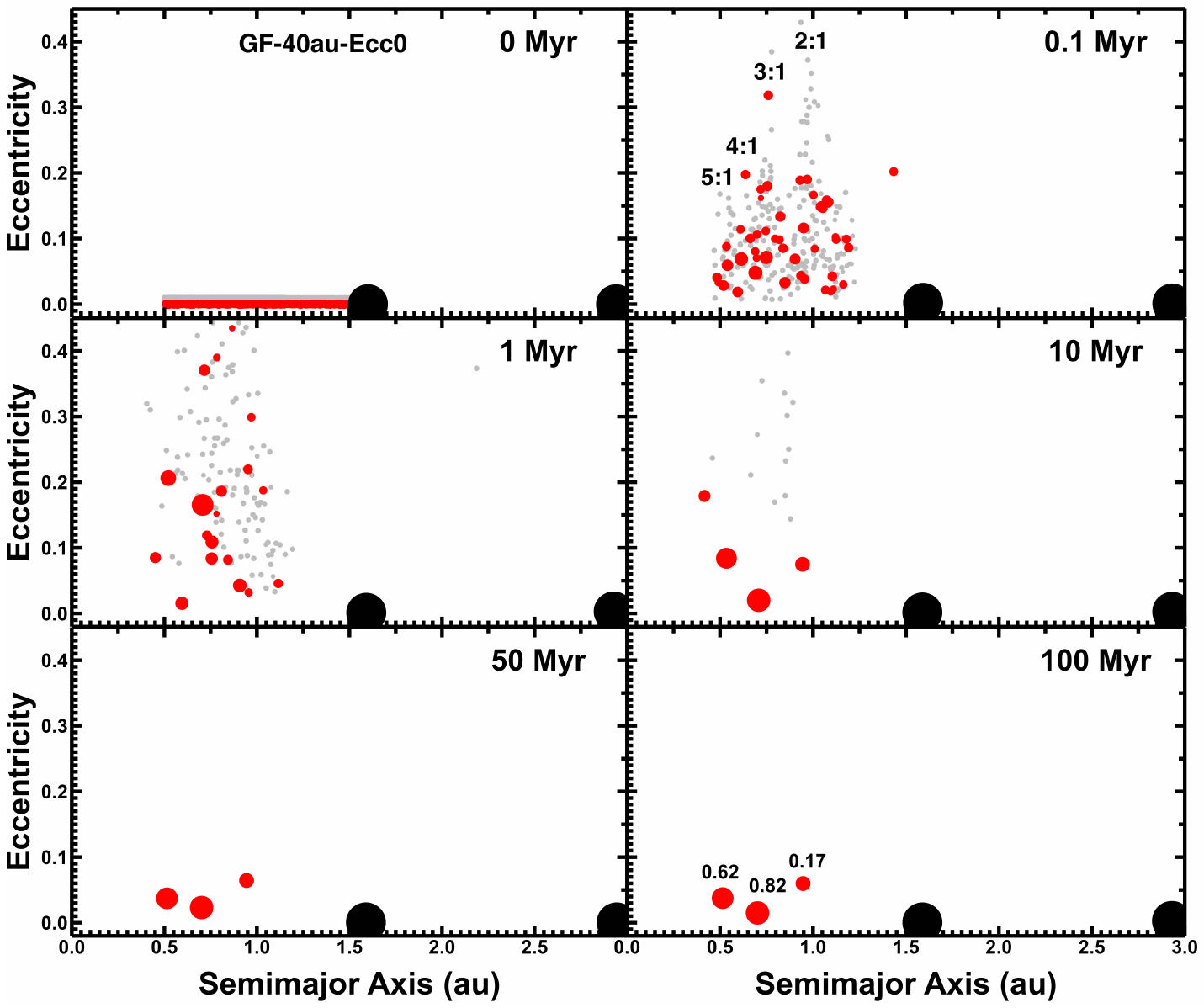}
\caption{Snapshots of planet formation simulations in a 40 au circular and coplanar binary for different secondary stars. 
The red circles represent planetary embryos and gray dots are planetesimals. The two giant planets, shown by black circles, 
are at 1.6 and 2.94 au. The figure shows the locations of some of the mean-motion resonances with the inner giant
planet. See Table 4 for the exact locations of these resonances. The secular resonances $({g_1} , {g_2})$ are at 
(0.35 , 0.67) au. The number above each red circle in the $t=100$ Myr panels indicates the final mass of that 
object in Earth masses. As shown here, the secular resonance of the outer planet is deeply suppressed and shows no significant 
contribution to the formation process.}
\label{fig1}
\end{figure*}

An inspection of Table 1 points to several interesting and important results. 
First, as shown here, in the majority of the simulations, the final system contains only one planet. Two-planet systems 
are rare and (as expected) appear in binaries where the perturbation of the secondary star is not strong
(e.g., in binaries with larger periastrons and/or smaller secondaries). Table 1 also shows 
that 90\% of the final planets in single-planet systems, and all the large planets in the double-planet systems, have masses 
between $\sim$0.6 and 1.7 Earth masses. This can be seen, for instance, in all simulations of figure 1. This is a significant 
result that suggests that, if the growth of planetesimals to planetary embryos proceeds constructively 
around a star of a moderately close binary, terrestrial planet formation may also proceed constructively in that system. 
Finally, the fact that many of the final planets in Table 1 are of Mars mass is another important result. As shown by this table,
10\% of the final bodies in systems with one planet, and almost all small bodies in double-planet systems, have masses in 
the range of 1.3 -- 4.5 Mars masses.

\begin{deluxetable*}{ccccccccccccc}
\tablecaption{Number, mass, and orbital properties of the final planets interior to the inner giant planet. 
All systems are coplanar and their giant planets are initially at 1.6 and 2.94 au.
The quantities $({a_{\rm b}},{e_{\rm b}},{q_{\rm b}})$ represent the semimajor axis, eccentricity and periastron 
distance of the binary, $a_{g_1}$ and $a_{g_2}$ are the initial locations of the secular resonances of the 
inner and outer planets, $N$ is the number of the final planets formed interior to the inner giant planet, and
${m_{\rm p}} , {a_{\rm p}}, {e_{\rm p}},$ and $i_{\rm p}$ denote the mass, semimajor axis, eccentricity and inclination
of the final planets, respectively. The quantity $\Delta {a_{\rm p}}$ demonstrate the displacement in semimajor axis of 
the final planet with respect to the initial location of its seed embryo. The time $T_{\rm p}$ corresponds  to the 
last major (i.e., embryo-embryo) impact that resulted in the final growth of the planet. The entries shown by
triple dots denote lack of planet growth (i.e., all objects were scattered outside the integration region).
\label{chartable}}
\tablehead{
\colhead{Binary} & \colhead{$a_{\rm b}$} & \colhead{$e_{\rm b}$} & \colhead{$q_{\rm b}$} & \colhead{$a_{g_1}$} & \colhead{$a_{g_2}$} 
& \colhead{$N$} & \colhead{$m_{\rm p}$} & \colhead{$a_{\rm p}$} & \colhead{$\Delta {a_{\rm p}}$} & \colhead{$e_{\rm p}$} &
\colhead{$i_{\rm p}$} & \colhead{$T_{\rm p}$} \\   
\colhead{} & \colhead{(au)} & \colhead{} & \colhead{(au)} & \colhead{(au)} & \colhead{(au)} & \colhead{} &
\colhead{($m_E$)} & \colhead{(au)} & \colhead{(au)} & \colhead{} & \colhead{(deg)} & \colhead{(Myr)}  
} 
\startdata
GG & 20 & 0   & 20 & 0.70 & 1.05 & 1 & 1.56        & 0.45         & -0.59        & 0.08        & 6.15        & 14.43        \\
GG & 20 & 0.1 & 18 & 0.70 & 1.05 & 2 & 0.96 , 0.13 & 0.56 , 0.73  & -0.01 , 0.20 & 0.02 , 0.04 & 2.20 , 3.36 & 5.94 , 1.59  \\
GG & 20 & 0.2 & 16 & 0.70 & 1.05 & 1 & 0.76        & 0.54         & -0.18        & 0.09        & 2.61        & 3.37         \\
GG & 20 & 0.3 & 14 & 0.70 & 1.05 & 1 & 0.58        & 0.54         & 0.01         & 0.13        & 0.87        & 0.45         \\
GG & 30 & 0   & 30 & 0.47 & 0.78 & 1 & 1.32        & 0.59         & 0.01         & 0.07        & 2.72        & 7.40         \\
GG & 30 & 0.1 & 27 & 0.47 & 0.78 & 1 & 0.94        & 0.52         & -0.23        & 0.12        & 2.6         & 5.70         \\
GG & 30 & 0.2 & 24 & 0.47 & 0.78 & 1 & 1.09        & 0.60         & -0.02        & 0.06        & 1.14        & 5.20         \\
GG & 30 & 0.3 & 21 & 0.47 & 0.78 & 1 & 0.38        & 0.66         & 0.09         & 0.50        & 2.04        & 0.68         \\
GG & 30 & 0.4 & 18 & 0.47 & 0.78 & 1 & 0.23        & 0.56         & 0.04         & 0.36        & 1.21        & 0.64         \\
GG & 30 & 0.5 & 15 & 0.47 & 0.78 & 1 & 0.19        & 0.37         & -0.29        & 0.30        & 0.83        & 1.20         \\
GG & 40 & 0   & 40 & 0.35 & 0.67 & 1 & 1.40        & 0.63         & -0.06        & 0.03        & 3.03        & 5.00         \\
GG & 40 & 0.1 & 36 & 0.35 & 0.67 & 1 & 1.34        & 0.60         & -0.02        & 0.06        & 2.12        & 6.12         \\
GG & 40 & 0.2 & 32 & 0.35 & 0.67 & 1 & 1.30        & 0.57         & -0.12        & 0.06        & 8.29        & 17.11        \\
GG & 40 & 0.3 & 28 & 0.35 & 0.67 & 0 & \nodata     & \nodata      & \nodata      & \nodata     & \nodata     & \nodata      \\
GG & 40 & 0.4 & 24 & 0.35 & 0.67 & 0 & \nodata     & \nodata      & \nodata     & \nodata     & \nodata     & \nodata      \\
GG & 40 & 0.5 & 20 & 0.35 & 0.67 & 1 & 0.42        & 0.57         & 0.02         & 0.09        & 7.73        & 1.39         \\
GG & 40 & 0.6 & 16 & 0.35 & 0.67 & 0 & \nodata     & \nodata      & \nodata      & \nodata     & \nodata     & \nodata      \\
GG & 50 & 0   & 50 & 0.28 & 0.62 & 1 & 1.66        & 0.58         & 0.00         & 0.09        & 2.88        & 38.34        \\
GG & 50 & 0.1 & 45 & 0.28 & 0.62 & 1 & 1.57        & 0.61         & -0.03        & 0.04        & 3.40        & 9.89         \\
GG & 50 & 0.2 & 40 & 0.28 & 0.62 & 1 & 1.17        & 0.57         & -0.08        & 0.07        & 2.78        & 4.36         \\
GG & 50 & 0.3 & 35 & 0.28 & 0.62 & 1 & 1.01        & 0.57         & 0.00         & 0.06        & 1.55        & 1.38         \\
GG & 50 & 0.4 & 30 & 0.28 & 0.62 & 1 & 0.66        & 0.56         & 0.04         & 0.08        & 0.20        & 1.13         \\
GG & 50 & 0.5 & 25 & 0.28 & 0.62 & 1 & 0.46        & 0.51         & -0.14        & 0.08        & 3.24        & 4.40         \\
GG & 50 & 0.6 & 20 & 0.28 & 0.62 & 1 & 0.77        & 0.57         & 0.00         & 0.22        & 0.90        & 3.15         \\
GG & 50 & 0.7 & 15 & 0.28 & 0.62 & 0 & 0.76        & 0.54         & -0.01        & 0.12        & 0.40        & 0.96      \\
GM & 20 & 0   & 20 & 0.53 & 0.84 & 1 & 1.69        & 0.58         & -0.07        & 0.05        & 4.66        & 9.50         \\
GM & 20 & 0.2 & 16 & 0.53 & 0.84 & 1 & 0.60        & 0.47         & -0.14        & 0.07        & 1.89        & 4.00         \\
GM & 40 & 0   & 40 & 0.35 & 0.67 & 2 & 0.44 , 1.23 & 0.46 , 0.69  & -0.26 , -0.22& 0.05 , 0.08 & 4.83 , 5.40 & 3.40 , 8.90  \\
GM & 40 & 0.2 & 32 & 0.35 & 0.67 & 2 & 0.92 , 0.36 & 0.53 , 0.81  & -0.22 , 0.29 & 0.16 , 0.06 & 1.03 , 2.64 & 3.30 , 1.06  \\
GM & 40 & 0.4 & 24 & 0.35 & 0.67 & 1 & 1.04        & 0.57         & -0.08        & 0.09        & 0.91        & 3.60         \\
GK & 20 & 0   & 20 & 0.63 & 0.94 & 2 & 0.44 , 1.13 & 0.51 , 0.69  & -0.06 , -0.40& 0.04 , 0.02 & 2.52 , 2.37 & 11.25 , 5.25 \\
GK & 20 & 0.2 & 16 & 0.63 & 0.94 & 1 & 0.80        & 0.54         & 0.01         & 0.12        & 1.22        & 2.46         \\
GK & 40 & 0   & 40 & 0.31 & 0.65 & 2 & 1.24 , 0.45 & 0.57 , 0.80  & 0.00 , -0.08 & 0.04 , 0.07 & 5.31 , 6.72 & 8.34 , 4.86  \\
GK & 40 & 0.2 & 32 & 0.31 & 0.65 & 2 & 0.59 , 0.25 & 0.49 , 0.97  & -0.15 , 0.34 & 0.06 , 0.10 & 1.88 , 2.17 & 2.72 , 0.26  \\
GK & 40 & 0.4 & 24 & 0.31 & 0.65 & 1 & 0.94        & 0.57         & 0.02         & 0.18        & 1.37        & 1.86         \\
GF & 20 & 0   & 20 & 0.76 & 1.06 & 1& 1.29        & 0.61         & -0.11        & 0.04        & 1.17        & 19.8         \\
GF & 20 & 0.2 & 16 & 0.76 & 1.06 & 1 & 1.15        & 0.61         & 0.06         & 0.06        & 1.25        & 8.28         \\
GF & 40 & 0   & 40 & 0.38 & 0.70 & 3 & 0.62 , 0.82 & 0.51 , 0.70  & 0.00 , -0.14 & 0.04 , 0.02 & 2.77 , 1.86 & 11.50 , 2.60 \\
   &    &     &    &      &      &   &     0.17    &     0.95     &     0.12     &     0.06    &     1.86    &      6.68    \\   
GF & 40 & 0.2 & 32 & 0.38 & 0.70 & 1 & 1.09        & 0.60         & -0.04        & 0.01        & 1.32        & 9.14         \\
GF & 40 & 0.4 & 24 & 0.38 & 0.70 & 1 & 0.61        & 0.52         & -0.07        & 0.23        & 1.81        & 1.74         \\
\enddata
\end{deluxetable*}

\begin{figure}[ht]
\vskip -7pt
\hskip -14pt
\hskip -0.1in
\includegraphics[scale=0.35]{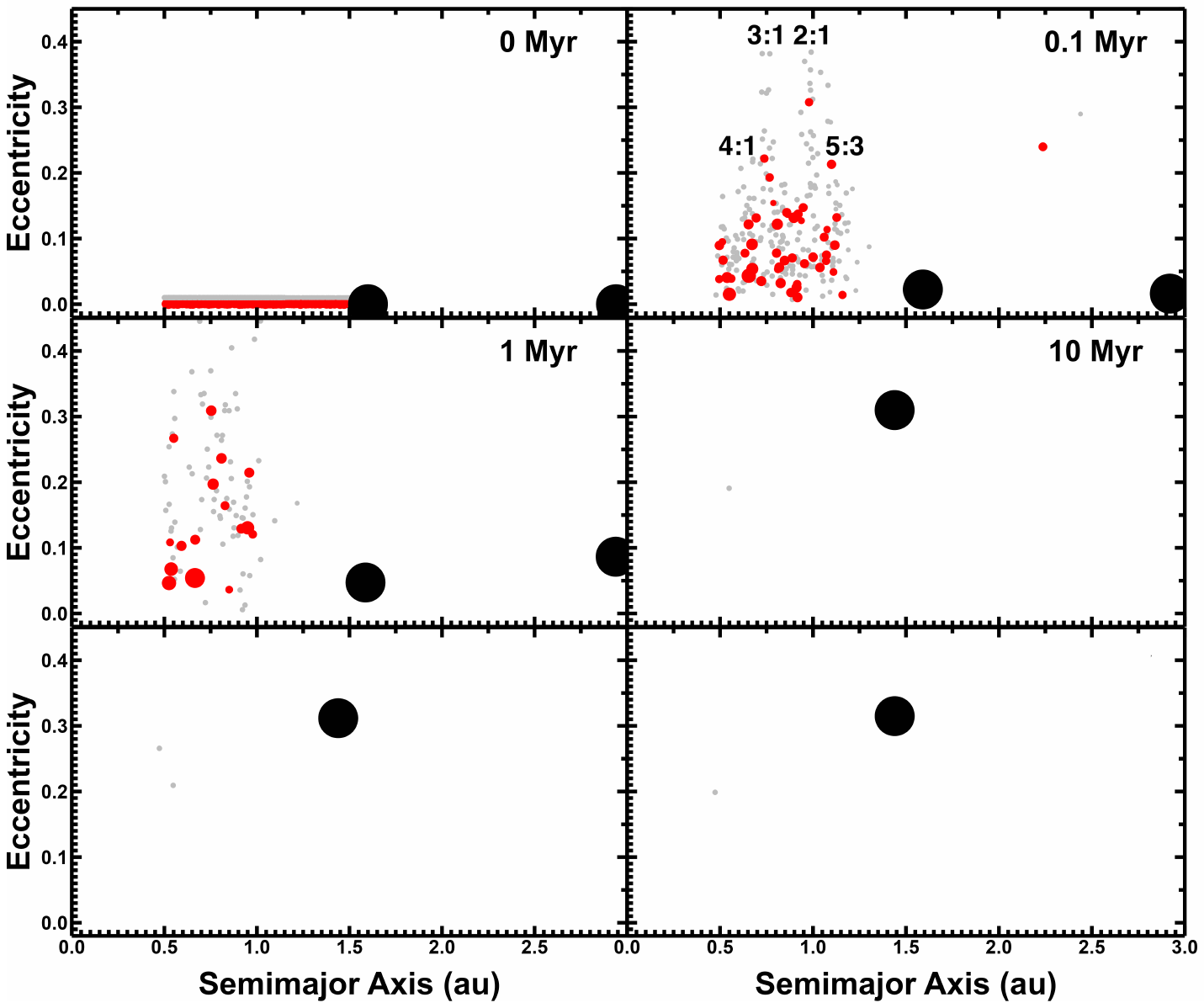}
\vskip -8pt
\caption{Snapshots of planet formation simulations in a 40 au coplanar binary with an eccentricity of 0.3.
The two giant planets, shown by black circles, are initially at 1.6 and 2.94 au. The figure shows the locations of 
some of the mean-motion resonances with the inner giant planet. As shown here, within the first 10 Myr, the outer planet 
is ejected from the system and the inner planet is scattered inward into an eccentric orbit. Note that as a result
of the strong interaction of the inner giant planet with the protoplanetary disk, all this material is scattered out
of the system and no planet is formed.}
\label{fig2}
\end{figure}

\begin{deluxetable}{cccccccc}
\tablecaption{The displacement in the semimajor axis of the inner and outer planets $(\Delta {a_{\rm {I/O}}})$, and their final 
eccentricities $({e_{\rm {I/O}}})$ at the end of the simulations of Table 1. In all the
simulations presented here, planets were initially in a circular and coplanar orbit with ${a_{\rm I}} = 1.60$ au and 
${a_{\rm O}} = 2.94$ au. Simulations with no entries denote cases where the outer planet was ejected.
\label{chartable}}
\tablehead{
\colhead{Binary} & \colhead{$a_{\rm b}$} & \colhead{$e_{\rm b}$} & \colhead{$q_{\rm b}$} & \colhead{$\Delta a_{\rm I}$} & \colhead{$\Delta a_{\rm O}$} 
& \colhead{$e_{\rm I}$} & \colhead{$e_{\rm O}$} \\   
\colhead{} & \colhead{(au)} & \colhead{} & \colhead{(au)} & \colhead{(au)} & \colhead{(au)} & \colhead{} & \colhead{} 
} 
\startdata
GG & 20 & 0   & 20 & -0.007 & -0.039 & 0.01 & 0.01        \\
GG & 20 & 0.1 & 18 & -0.015 & -0.021 & 0.01 & 0.04        \\
GG & 20 & 0.2 & 16 & -0.089 & 0.057  & 0.11 & 0.18        \\
GG & 20 & 0.3 & 14 & -0.095 & \nodata & 0.07 & \nodata    \\
GG & 30 & 0   & 30 & -0.012 & -0.023 & 0.00 & 0.00       \\
GG & 30 & 0.1 & 27 & -0.016 & -0.028 & 0.00 & 0.02        \\
GG & 30 & 0.2 & 24 & -0.015 & -0.033 & 0.02 & 0.01        \\
GG & 30 & 0.3 & 21 & -0.149 & \nodata & 0.06 & \nodata        \\
GG & 30 & 0.4 & 18 & -0.141 & \nodata & 0.11 & \nodata       \\
GG & 30 & 0.5 & 15 & 0.121 & \nodata & 0.04 & \nodata       \\
GG & 40 & 0   & 40 & -0.014 & 0.001 & 0.00 & 0.00        \\
GG & 40 & 0.1 & 36 & -0.014 & -0.003 & 0.00 & 0.00        \\
GG & 40 & 0.2 & 32 & -0.014 & -0.018 & 0.01 & 0.02        \\
GG & 40 & 0.3 & 28 & -0.161 & \nodata & 0.31 & \nodata     \\
GG & 40 & 0.4 & 24 & -0.020 & -0.015 & 0.03 & 0.08     \\
GG & 40 & 0.5 & 20 & -0.135 & \nodata & 0.09 & \nodata      \\
GG & 40 & 0.6 & 16 & -0.096 & -2.673 & 0.54 & 0.23     \\
GG & 50 & 0   & 50 & -0.012 & -0.013 & 0.00 & 0.01        \\
GG & 50 & 0.1 & 45 & -0.012 & 0.003 & 0.00 & 0.00        \\
GG & 50 & 0.2 & 40 & -0.017 & 0.007 & 0.00 & 0.01       \\
GG & 50 & 0.3 & 35 & -0.016 & -0.009 & 0.01 & 0.02       \\
GG & 50 & 0.4 & 30 & -0.021 & 0.013 & 0.03 & 0.11        \\
GG & 50 & 0.5 & 25 & -0.019 & -0.007 & 0.01 & 0.04        \\
GG & 50 & 0.6 & 20 & -0.134 & \nodata & 0.19 & \nodata        \\
GG & 50 & 0.7 & 15 & 0.175 & \nodata & 0.04 & \nodata     \\
GM & 20 & 0   & 20 & -0.010 & -0.021 & 0.00 & 0.01        \\
GM & 20 & 0.2 & 16 & -0.015 & -0.038 & 0.02 & 0.02        \\
GM & 40 & 0   & 40 & -0.014 & 0.011 & 0.00 & 0.00       \\
GM & 40 & 0.2 & 32 & -0.015 & 0.001 & 0.01 & 0.01        \\
GM & 40 & 0.4 & 24 & -0.017 & 0.010 & 0.02 & 0.02        \\
GK & 20 & 0   & 20 & -0.011 & -0.030 & 0.00 & 0.00        \\
GK & 20 & 0.2 & 16 & -0.015 & -0.056 & 0.01 & 0.03        \\
GK & 40 & 0   & 40 & -0.013 & -0.016 & 0.00 & 0.00        \\
GK & 40 & 0.2 & 32 & -0.013 & 0.001 & 0.02 & 0.01        \\
GK & 40 & 0.4 & 24 & -0.018 & 0.006 & 0.03 & 0.03        \\
GF & 20 & 0   & 20 & -0.008 & -0.036 & 0.00 & 0.01         \\
GF & 20 & 0.2 & 16 & -0.059 & \nodata & 0.03 & \nodata        \\
GF & 40 & 0   & 40 & -0.011 & -0.009 & 0.00 & 0.00        \\
GF & 40 & 0.2 & 32 & -0.014 & -0.002 & 0.01 & 0.02        \\
GF & 40 & 0.4 & 24 & -0.015 & -0.026 & 0.02 & 0.05       \\
\enddata
\end{deluxetable}

\begin{figure*}[ht]
\hskip -0.43in
\includegraphics[scale=0.7]{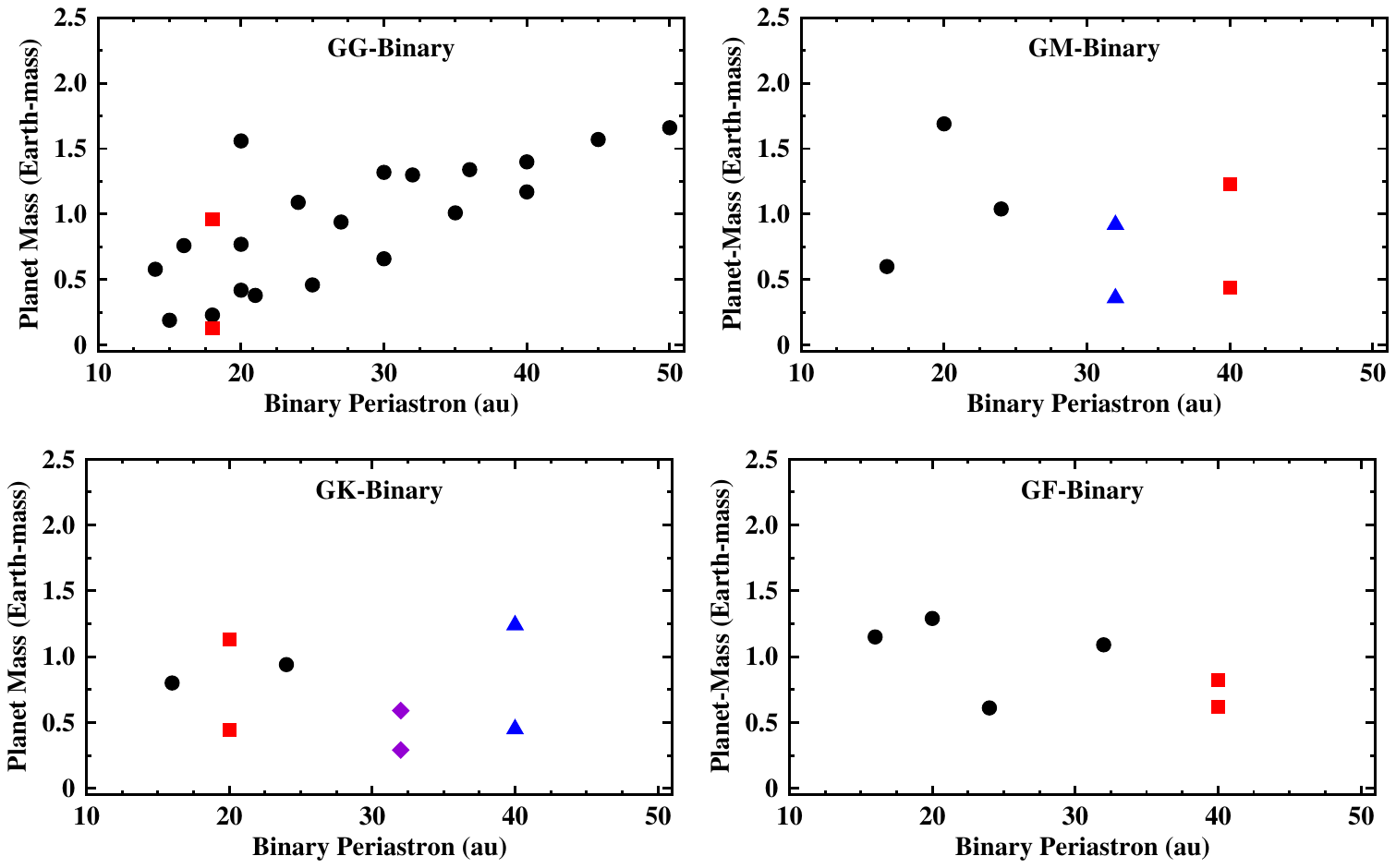}
\vskip -1.25in
\caption{Graphs of the masses of the final terrestrial planets in term of the binary periastron for all simulations of Table 1. Black circles
represent cases where the simulation produced only one planet. Colored symbols correspond to double-planet systems. Each two symbols with 
the same color formed in the same simulation. Although these simulations are stochastic, a trend can be observed in the panel for the GG 
binary where the masses of the final planets become larger for binaries with larger periastron distances. See the text for more details.}
\label{fig3}
\end{figure*}

\subsection{The effect of the secondary star}

While it is understood that the above findings are results of the dynamical evolution of the protoplanetary disk, and 
that the disk is strongly affected by the perturbation of the giant planets, it is important to recognize that, because the entire 
system revolves around the primary star, how the perturbation of the giant planets affects the dynamics of the protoplanetary 
material depends on how the orbits of these planets have been affected by the perturbation of the secondary. 
In fact, as demonstrated by \citet{Haghighipour07}, the giant planets play the important role of transferring the secondary's perturbation 
to the bodies interior to them, meaning that the strength and extent of their perturbing effects vary with the 
characteristics of the binary. Table 2 shows this for all simulations of Table 1. Shown here are the displacements in the semimajor axes 
of the two giant planets, and their final eccentricities at the end of each simulation.
As demonstrated by the last column, in binaries with large eccentricities (i.e., small periastron distances), 
the eccentricity of the outer giant planet becomes large as well. In several of these systems, the perturbation
of the secondary star becomes so strong that it ejects the outer planet out of its orbit. Figure 2 shows an example of such
systems. The binary in this figure has a semimajor axis of 40 au and its eccentricity is 0.3. The two giant planets are
initially at 1.6 and 2.94 au. The figure shows some of the mean-motion resonances with the inner giant planet as well.
As shown here, within the first 10 Myr, the outer giant planet is ejected and the inner planet is scattered
slightly inward into an orbit with an eccentricity of $\sim 0.3$. The strong interaction of this planet with the protoplanetary material
has scattered all these objects out of the system, resulting in no planet formation in the inner regions. Table 2 shows more examples of 
such systems. An interesting case in this table is that of the GG binary with the semimajor axis of 40 au and  eccentricity of 0.6.
As shown here, the outer planet is scattered inward to 0.27 au, most likely on its path to collide with the central star.

\begin{figure}[ht]
\vskip -9pt
\hskip -17pt
\includegraphics[scale=0.37]{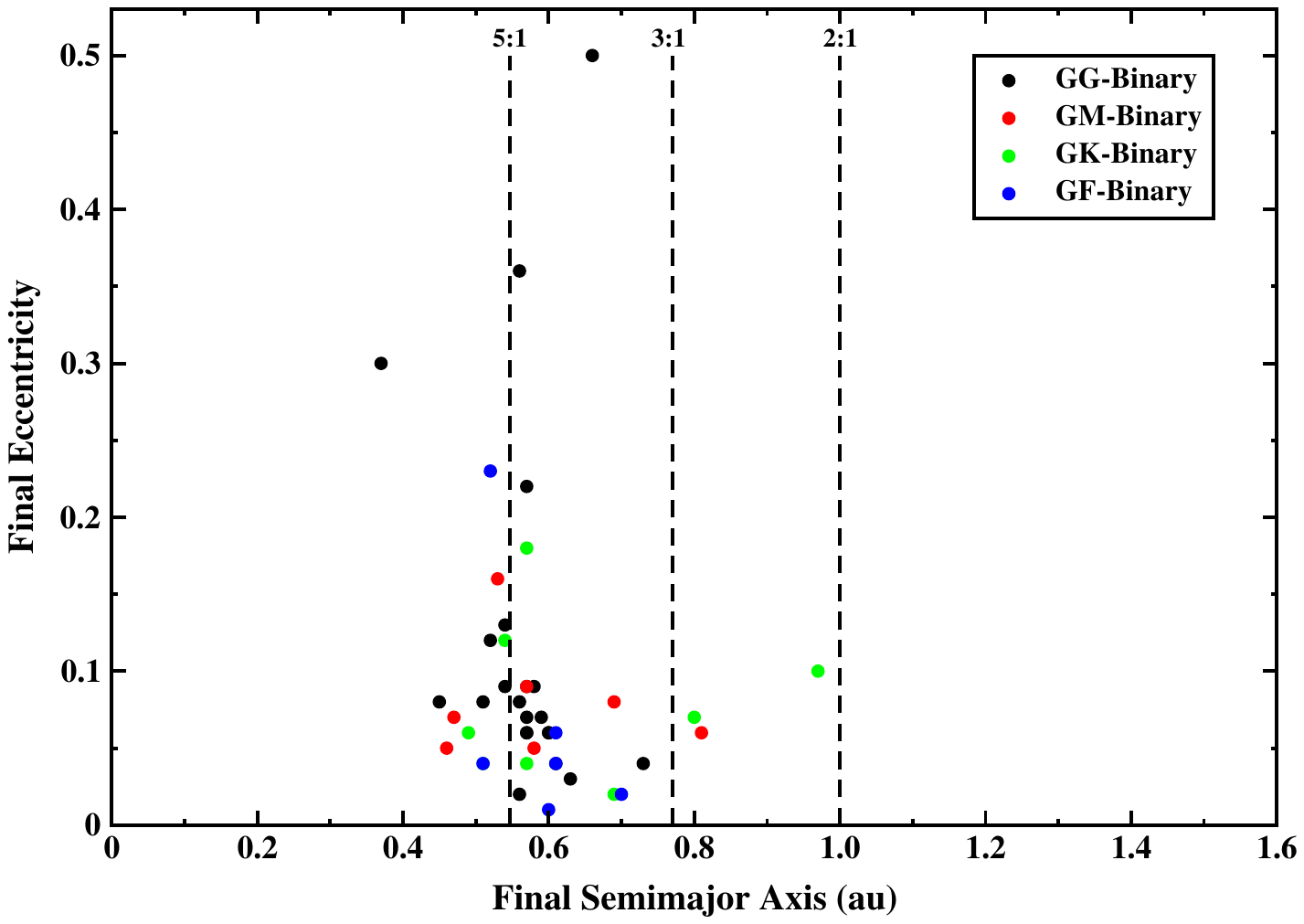}
\vskip -11pt
\caption{Graph of the semimajor axis and eccentricity of the final terrestrial planets in the simulations of Table 1. Note that the 
majority of the planets formed or reached their final orbits in the range of 0.45 -- 0.65 au.}
\label{fig4}
\end{figure}

The fact that the perturbation of the secondary star increases the eccentricity of the outer giant planet has immediate implications
for the masses and orbital characteristics of the final bodies. As demonstrated above, 
an increase in the eccentricity of this planet intensifies its perturbing
effect on the inner giant planet, which in turn increases the orbital eccentricity of this body, enhancing its close approaches to the 
protoplanetary disk. These close approaches remove planet-forming material by scattering the disk's objects out of the system or causing them 
to collide with the giant planets and central star (see Figure 9 in Paper I). The subsequent lack of planet-forming material results in 
the formation of fewer and/or smaller planets, or no planet at all.  We would like to emphasize that due to the stochasticity of the 
simulations, results shown in Table 2 must be taken qualitatively.

In contrast, in binaries where the perturbation of the secondary star is not strong, the disk can maintain more of its material for planet 
formation and growth. In general, this availability of more material manifests itself in two ways: either the system forms one large 
planet as in most of the single-planet systems, or it forms multiple planets with smaller masses as in all double-planet systems. 
Figure 3 shows this for all simulations of Table 1. In this figure, we show the masses of the final planets in terms of the 
periastron distances of their host binaries. Black circles represent the final bodies in single-planet systems, and the colored 
symbols are the final bodies in systems that formed two planets. Each two planets of the same color formed in the same simulation. 
The top left panel shows the effect of the binary periastron in an equal-mass binary. As shown here, planet masses are small for binaries 
with small periastrons, and they increase as this quantity becomes larger. When the secondary star is replaced with a more massive one 
(e.g., bottom right panel for an F star), the increase in its perturbation causes the planets in binaries with same periastron 
distances to become smaller (compare the masses of the planets at ${q_{\rm b}}=20$ au and ${e_{\rm b}}=0$ in the GG and GF binaries). 
However, when the mass of the secondary is small, the formation process tends to form pairs of planets; see the GM and GK panels. These 
panels also show that when the binary periastron is large, not only do the final planets come in a pair, they carry more mass as well.

As mentioned before, in almost all double-planet system, the larger planet is Earth mass and the smaller planet has a mass of 
a few times that of Mars. This connection between the growth and masses of these two planets resembles the connection between the 
growth of Earth and the formation of Venus. As mentioned in the introduction, in the traditional simulations of terrestrial 
planet formation in our solar system, where Jupiter and Saturn are assumed to be in their current orbits, the secular resonance 
of Saturn confines the inner parts of the protoplanetary disk to the region interior to 2.1 au. Simulations of the late stage
of terrestrial planet formation in such systems have demonstrated that in the majority of the cases, the final system 
contains an Earth analog plus a second planet with a slightly smaller mass in an inner orbit, i.e., a Venus analog 
\citep{Haghighipour16}. In these simulations, Earth appears as an organic product of the evolution of the 
protoplanetary disk, and Venus comes as a bonus. Our simulations point to a similar trend, with our
Earth analogs being the natural outcome of the evolution of the system, and in those cases where the system manages to 
maintain enough material for forming a second planet, that planet will be in an exterior orbit and most likely a Mars analog

\subsection{Effects of the Secular and Mean-motion Resonances}

To explore the effects of the resonances, we examined the orbital evolution of the final planets during their formation and
after they were fully formed. Figure 4 shows the semimajor axes and eccentricities of these bodies at the end of the simulations
of Table 1. As shown here, although the initial protoplanetary disk extended from 0.5 to 1.5 au, the large majority of the planets 
settled in orbits with semimajor axes between 0.45 and 0.65 au. Compared to the locations of secular resonances in Table 1, 
in almost all cases, these orbits are interior to the secular resonance of the outer planet, $g_2$. While this seems to be consistent 
with the effect of the secular resonance of Saturn in our solar system (recall that in our solar system, the secular resonance of Saturn, 
$\nu_6$, confined the terrestrial planet region to inside 2.1 au), an examination of the quantity $\Delta a$, the displacement of a planet 
during its formation, indicates that most of these bodies migrated inward and passed through the location of 
$g_2$ without being altered by this resonance. This is a significant result that indicates that the secular resonance of the outer 
giant planet did not have a significant effect on the evolution of the protoplanetary disk and the growth of the final 
bodies. In other words, this resonance was strongly suppressed. 

To examine the effect of the secular resonance of the inner giant planet, $g_1$, we note that as shown by Table 1, in the majority
of the systems, the location of this resonance falls interior to the inner edge of the protoplanetary disk (0.5 au). In those cases 
where this resonance was within the disk, either the final planets did not pass through it, or those few that did, were not 
affected by it and continued their growth uninterruptedly. This is not an unexpected result 
as similar process occurs in our solar system as well: the secular resonance of Jupiter at $\sim 0.9$ au is heavily suppressed by the 
perturbation of Saturn and hardly disturbs planet formation in and around its location.

\begin{deluxetable*}{cccccccccccc}
\tablecaption{Similar to Table 1, but for different values of the semimajor axes of the giant planets $({a_{\rm I} }\,,{a_{\rm O}})$ in a coplanar
GG binary with a semimajor axis of 20 and 30 (au) and eccentricity of 0 and 0.2.
\label{chartable}}
\tablehead{
\colhead{$a_{\rm b}$} & \colhead{$e_{\rm b}$} & \colhead{$q_{\rm b}$} & \colhead{$a_{\rm I}$} & \colhead{$a_{\rm O}$} 
& \colhead{$N$} & \colhead{$m_{\rm p}$} & \colhead{$a_{\rm p}$} & \colhead{$\Delta {a_{\rm p}}$} & \colhead{$e_{\rm p}$} &
\colhead{$i_{\rm p}$} & \colhead{$T_{\rm p}$} \\   
\colhead{(au)} & \colhead{} & \colhead{(au)} & \colhead{(au)} & \colhead{(au)} & \colhead{} &
\colhead{($m_E$)} & \colhead{(au)} & \colhead{(au)} & \colhead{} & \colhead{(deg)} & \colhead{(Myr)}  
} 
\startdata
20 & 0   & 20 & 1.36 & 2.50 & 1 & 0.93        & 0.50         & -0.06        & 0.18        & 4.66        & 2.94        \\
20 & 0   & 20 & 1.60 & 2.94 & 1 & 1.56        & 0.45         & -0.59        & 0.08        & 6.15        & 14.43        \\
20 & 0   & 20 & 1.90 & 3.50 & 3 & 0.10 , 0.64 & 0.43 , 0.57  & -0.16 , -0.22 & 0.08 , 0.07 &  4.12 , 1.02 & 13.7 , 4.80  \\
   &     &   &       &      &   & 0.96         & 0.78        & -0.05          & 0.04         & 0.44        & 16.7        \\
20 & 0   & 20 & 2.17 & 4.00 & 2 & 1.11 , 0.99  & 0.57 , 0.94 & 0.00 , 0.20  & 0.07 , 0.04  & 2.52 , 0.98  & 18.00  , 9.55  \\
30 & 0   & 30 & 1.36 & 2.50 & 1 & 0.99        & 0.55         & -0.14        & 0.07        & 2.45        & 5.95        \\
30 & 0   & 30 & 1.60 & 2.94 & 1 & 1.31        & 0.59         & 0.01         & 0.07        & 2.72        & 7.40       \\
30 & 0   & 30 & 1.90 & 3.50 & 3 & 0.87 , 0.65  & 0.56 , 0.73 & -0.23 , 0.16 & 0.03 , 0.04  & 1.27 , 5.30 & 6.68 , 10.9 \\
   &     &   &       &      &   & 0.19         & 1.06        & 0.02          & 0.12         & 8.95        & 5.25        \\
30 & 0   & 30 & 2.17 & 4.00 & 2 & 0.93 , 1.11  & 0.56 , 0.82 & -0.42 , 0.17  & 0.04 , 0.03  & 2.85 , 2.37 & 41.95  , 6.33  \\
20 & 0.2 & 16 & 1.36 & 2.50 & 2 & 0.43 , 0.18  & 0.48 , 0.70  & -0.21 , -0.06  & 0.05 , 0.05  & 0.60 , 2.13  & 1.28 , 2.94   \\
20 & 0.2 & 16 & 1.60 & 2.94 & 1 & 0.76         & 0.54         & -0.18          & 0.09        & 2.61         & 3.37         \\
20 & 0.2 & 16 & 1.90 & 3.50 & 2 & 0.50 , 0.47  & 0.55 , 0.70 & 0.03 , 0.07 & 0.04 , 0.05 & 0.60 , 0.30 & 2.94 , 0.75        \\
20 & 0.2 & 16 & 2.17 & 4.00 & 2 & 0.44 , 1.25  & 0.56 , 0.61  & -0.18 , -0.08  & 0.31 , 0.41   & 6.65 , 5.00  & 3.08 , 4.10         \\
30 & 0.2 & 24 & 1.36 & 2.50 & 1 & 0.52        & 0.50         & -0.02        & 0.05        & 2.47        & 1.80         \\
30 & 0.2 & 24 & 1.60 & 2.94 & 1 & 1.09        & 0.60         & -0.02        & 0.06        & 1.14        & 5.20        \\
30 & 0.2 & 24 & 1.90 & 3.50 & 1 & 1.09        & 0.56         & 0.04        & 0.01        & 4.40        & 6.44         \\
30 & 0.2 & 24 & 2.17 & 4.00 & 2 & 1.43 , 0.20 & 0.65 , 1.06  & -0.07 , -0.03  & 0.03 , 0.16  & 0.40 , 3.10  & 6.00 , 14.00         \\
\enddata
\end{deluxetable*}

While the secular resonances did not play a significant role in the evolution of the disk and the process of terrestrial planet 
formation, the mean-motion resonances of the inner giant planet showed strong contributions to shaping the accretion region and
the mass and orbital architecture of the final bodies. As shown by Figure 1 (and, as we will demonstrate in the next sections,
by Figures 5 and 6), these resonances greatly disturb the evolution of the protoplanetary disk, causing many of the disk's objects 
to be scattered out of the system. This scattering effect is much stronger in the regions of low-order resonances, covering a 
large area interior to the inner giant planet. The latter confines planet formation to interior to $\sim 0.65$ au
where the orbits of the large majority of the Earth analogs are (the accumulation of orbits at 5:1 in Figure 4 is due to 
the fact that in the simulations of Table 1, this resonance is located close to the edge of the disk at 0.55 au). It also 
causes the outer region of the disk to be void of material for
planet formation (hence the formation of single-planet systems), or to maintain so small an amount that only Mars-mass
planets can form (hence the appearance of Mars analogs in the outer part of the disk). As demonstrated in Sections 3.3 and
3.4, these effects are enhanced when the semimajor axes and inclinations of the two planets are increased.

\begin{figure*}[ht]
\center
\vskip 12pt
\hskip -14pt
\includegraphics[scale=0.35 , angle=-90]{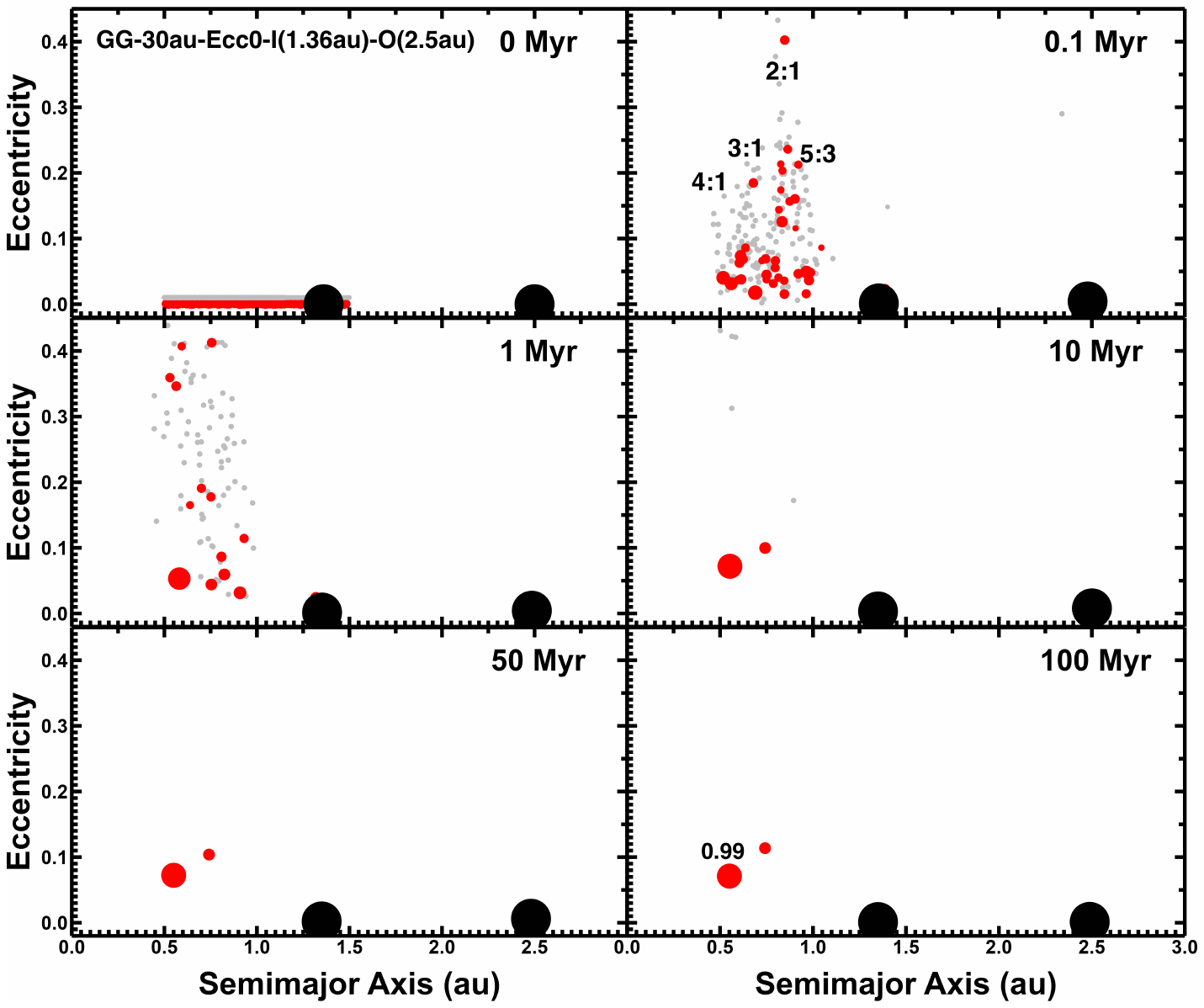}
\hskip -33pt
\includegraphics[scale=0.35 , angle=-90]{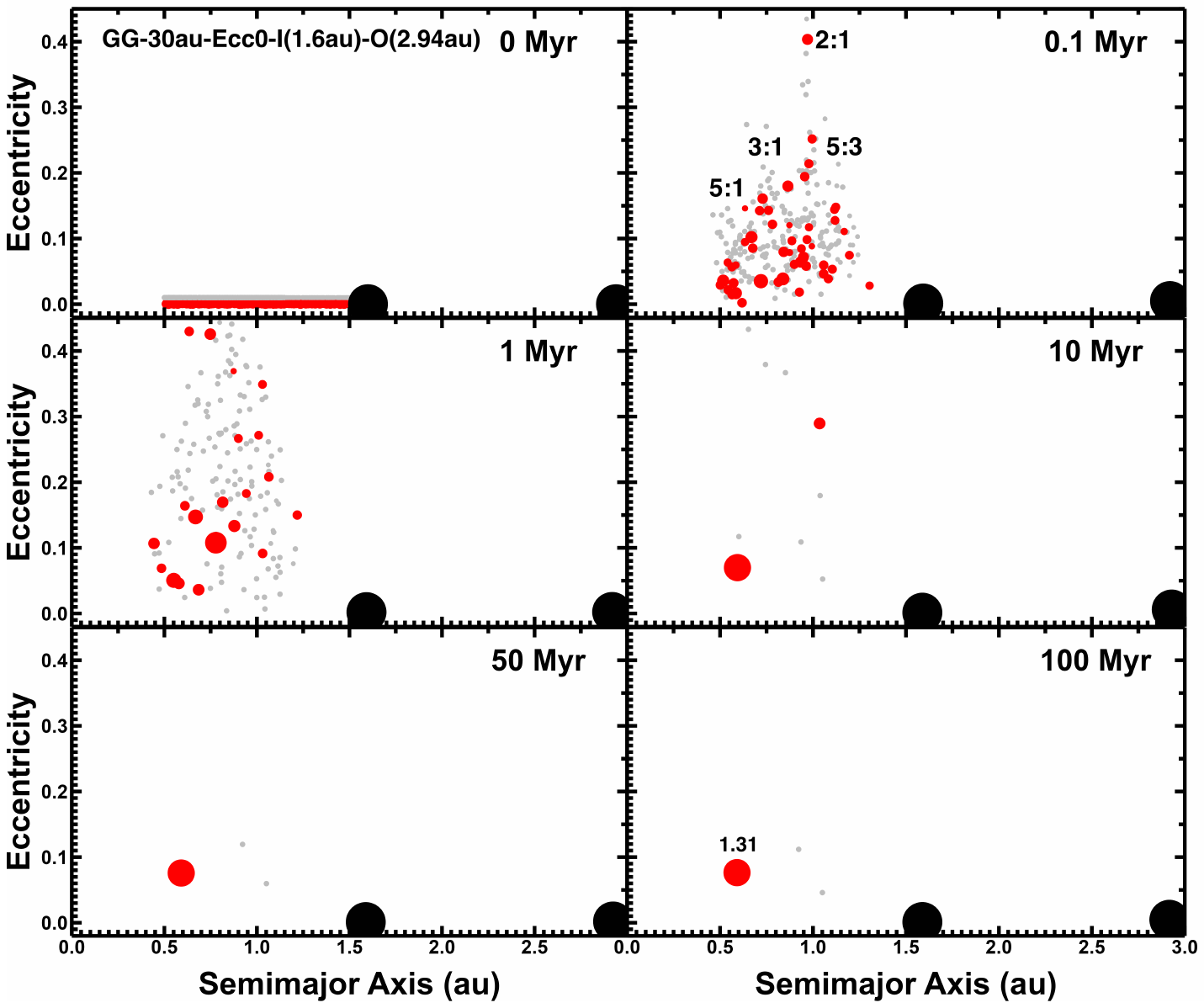}
\vskip -5pt
\hskip -14pt
\includegraphics[scale=0.35 , angle=-90]{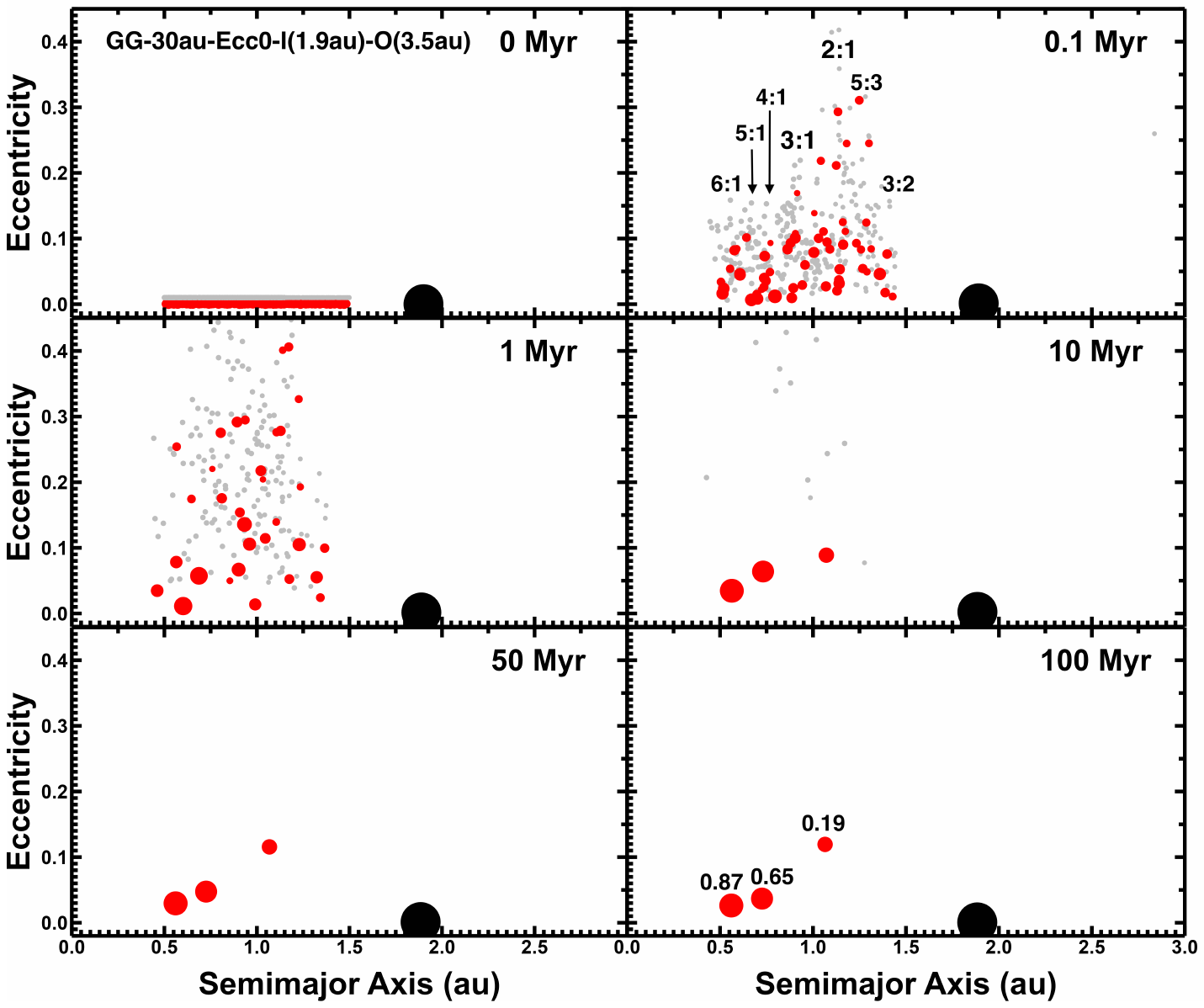}
\hskip -33pt
\includegraphics[scale=0.35 , angle=-90]{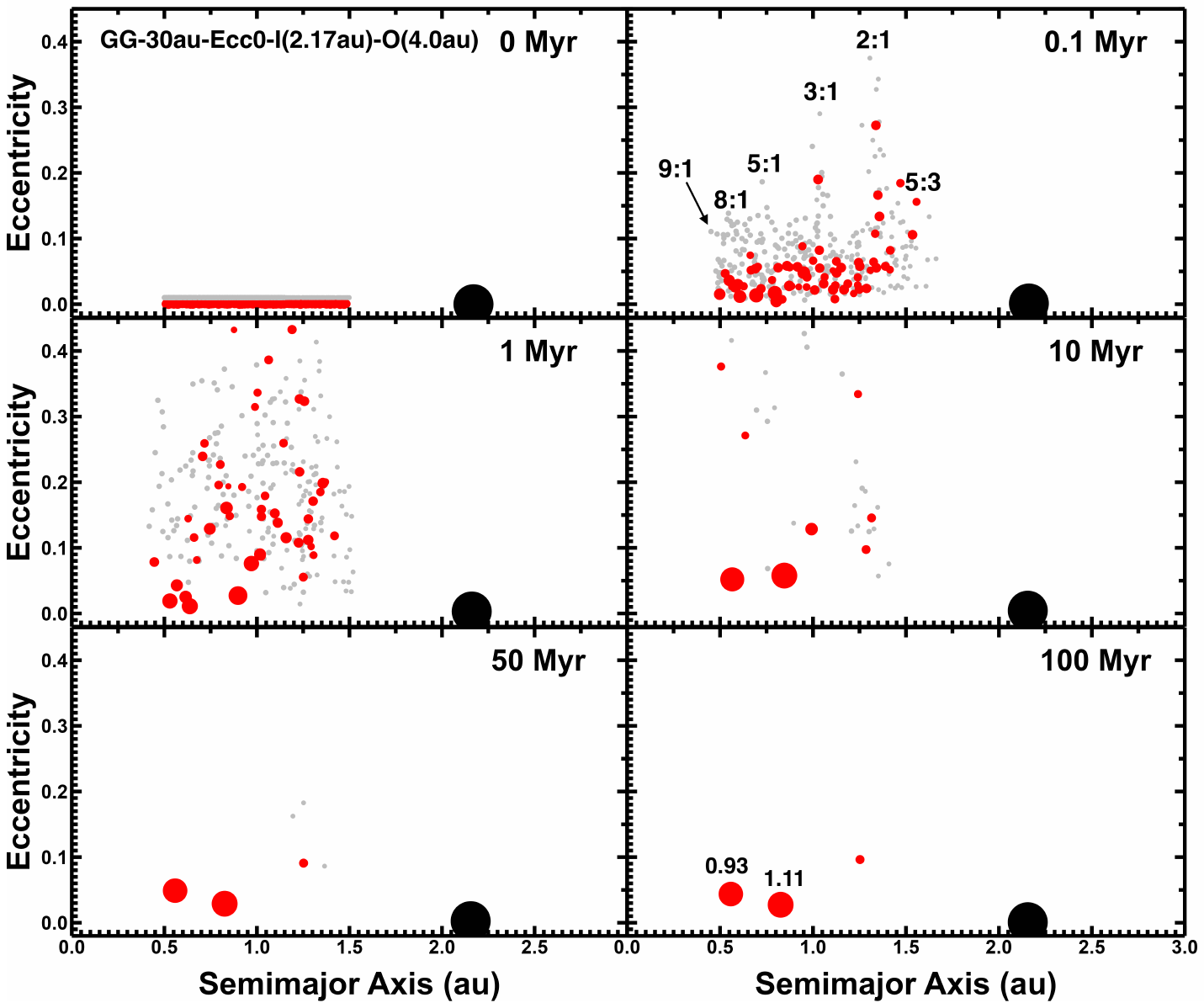}
\caption{Snapshots of the planet formation simulations for different values of the semimajor axes of the giant planets 
in a circular and coplanar 30 au GG binary. The number above each red circle in the $t=100$ 
Myr panel indicates the final mass of that object in Earth mass. Note that in the naming of each simulation, 
I stands for the inner planet, O is the outer planet, and the numbers in parentheses correspond to their initial semimajor 
axes. In order to be consistent with Figure 1, and to be able to show some of the structures of the protoplanetary disk, 
particularly, the appearance of the additional mean-motion resonances for larger values of the semimajor axis of the inner giant planet, 
we maintained the range of the semimajor axis at 3.0 au. As a result, the outer giant planet does not appear in the two lower 
simulations as it falls outside this range.}
\label{fig5}
\end{figure*}

\subsection{Varying Giant Planets' Semimajor Axes}

As mentioned earlier, we carried out simulations for four different values of the semimajor axis of the outer planet:
2.5, 2.94, 3.5 and 4.0 (au). At the start of each simulation, the inner planet was placed in an orbit with the
same period-commensurability as that of Jupiter and Saturn. Table 3 shows the results for a GG binary 
with a semimajor axis of 20 and 30 (au), and an eccentricity of 0 and 0.2. As expected, results pointed to similar 
trends to those in Table 1 with one additional feature: the sum of the planetary masses at the end of the
simulations increased by increasing the giant planets' semimajor axes. Figure 5 shows this in more detail. 
We show in this figure the evolution of the protoplanetary disk for all combinations of the semimajor axes of the two 
giant planets in a circular and coplanar, 30 au GG binary. As shown here, the disk is greatly altered by the mean-motion 
resonances of the inner giant planet. Table 4 shows the locations of these resonances. As demonstrated by this table and the 
panels at $t=0.1$ Myr, when the inner giant planet is initially at a larger semimajor axis, more resonances appear in the 
region of 0.5 -- 1.5 au. While in the outer part of the disk, the interaction of the disk's objects with these resonances 
scatters many of them out of the system, in the inner part, where the resonances are of high orders and their
scattering effects are not as strong, the interaction of disk material with these resonances increases the eccentricities of these 
objects, which enhances the probability of their orbit-crossing and ultimately their collisional growth. The two bottom simulations 
in figure 5 show this more clearly. As shown by the panels at $t=0.1$ Myr, with the inner planet being farther away, more of 
the disk material becomes available and subject to larger number of (high-order) mean-motion resonances. As a result, 
both the number of the final planets and the total planetary mass increase in these simulations. We note that, as shown 
in Figure 5, the secular resonances of the giant planets at 0.47 and 0.78 au are strongly suppressed.

\begin{deluxetable}{ccccc}
\tablecaption{Locations of the mean-motion resonances of the inner giant planet.
\label{chartable}}
\tablehead{
\colhead{$a_{\rm I}$ (au)} & \colhead{1.36} & \colhead{1.60} & \colhead{1.90} & \colhead{2.17} \\
\colhead{$a_{\rm O}$ (au)} &  \colhead{2.50} & \colhead{2.94} &  \colhead{3.50} & \colhead{4.0} \\   
\colhead{MMR} & \colhead{(au)} & \colhead{(au)} & \colhead{(au)} & \colhead{(au)} 
} 
\startdata
3:2 & 1.04 & 1.22 & 1.45 & 1.66  \\
5:3 & 0.97 & 1.14 & 1.35 & 1.54 \\
2:1 & 0.86 & 1.00 & 1.19 & 1.37 \\
3:1 & 0.65 & 0.77 & 0.91 & 1.04 \\
4:1 & 0.54 & 0.63 & 0.75 & 0.86 \\
5:1 &      & 0.55 & 0.65 & 0.74 \\
6:1 &      &      & 0.57 & 0.66 \\
7:1 &      &      &      & 0.59 \\
8:1 &      &      &      & 0.54 \\
9:1 &      &      &      & 0.50 \\
\enddata
\end{deluxetable}

\subsection{Varying Giant Planets' Orbital Inclinations}

When developing the theoretical foundations of our project, we assumed in Paper I that the entire system consisting 
of the binary, giant planets,
and protoplanetary disk were coplanar. This assumption was based on the consideration that during planet formation, 
the dynamical friction due to the collisional debris and the bimodality in the mass of the solid objects (i.e., the population 
of small planetesimals versus Moon- to Mars-sized planetary embryos) damp the eccentricities and inclinations of the growing planets. 
However, as bodies grow and interact with one another, it is not uncommon (especially in multiplanet systems) for objects to 
deviate from the general coplanarity and maintain inclined orbits. This can be seen in column $i_{\rm p}$ in Table 1 where the 
orbits of the final planets carry small to modest inclinations. 

In systems where the orbits are inclined, the eccentricity and inclination decouple during the secular evolution of the system 
\citep[e.g.,][]{Li14}. This decoupling alters the extent to which the perturbation of the giant planets affects the dynamics 
of the protoplanetary disk and the outcome of planet formation. To examine this effect, we carried out a series of simulations 
where the orbital inclinations of the two giant planets were taken to be $5^\circ$ and $10^\circ$. To maintain focus
on the effect of the inclination, and in order to be able to compare the results with coplanar cases (Table 1), we considered a 
GG binary with a moderate semimajor axis of 30 au, and placed the giant planets at the semimajor axes of 1.60 and 2.94 au.  
We then integrated the system for two values of the binary eccentricity,  ${e_{\rm b}}=0$ and 0.2.

Tables 5 and 6 show the results. As shown here, in general, inclination has a negative effect. The planets in these systems are 
fewer and less massive than those in coplanar systems. Also, their orbits are more eccentric. This all can be 
attributed to the fact that the dynamics of the disk is primarily driven by the perturbation of the inner giant planet.
Figure 6 shows this in more detail. We show in this figure the results of the simulations for an inclined outer planet
(upper right), an inclined inner planet (bottom left), and when the orbits of both planets are inclined (bottom right). 
As a point of comparison, we also show the simulation when the system is coplanar (upper left). As shown here,
an inclined orbit for either of the planets strongly enhances the mean-motion resonances of the inner giant planet. 
These resonances are much stronger in systems where the orbit of the inner planet is inclined (the two bottom simulations).
The latter enhances the efficiency of the resonances in removing material, which causes fewer planets to form, and 
those that form will have smaller masses. The perturbing effect of an inclined inner giant planet is so strong that in a moderately 
eccentric binary, the disk loses all its material and no planet is formed. This can be seen in Table 6 where simulations were
carried out for a binary with an eccentricity of 0.2. As shown here, compared to Table 5, the addition of the binary eccentricity
caused the masses of the final planets to become smaller and their orbital eccentricities become larger. In the last three
simulations, this perturbation becomes so strong that the systems cannot form any planets.

Finally, a comparison between the semimajor axes of the final planets, the quantity $\Delta{a_{\rm p}}$ in Tables 5 and 6,
and the locations of the secular resonances in these systems $({a_{g_1}}=0.47 , {a_{g_2}}=0.78)$ (au) indicates 
that many of these planets migrated through the secular resonances without their orbits being altered or themselves being
scattered out of the system. Some of these planets even maintained 
stable orbits at the location of the inner secular resonance. These findings once again confirm the prediction of the general theory
(Paper I) that the secular resonances are suppressed when the system is subject to the perturbation of an additional star. This
suppressing effect of the secondary is so strong that it cannot be offset even when the inclination is decoupled from eccentricity 
in inclined systems.

\begin{deluxetable*}{ccccccccc}
\tablecaption{Number, mass, and orbital properties of the final terrestrial planets in a 30 au, circular, GG binary for
different values of the orbital inclinations of the giant planets. Quantities $i_{\rm I}$ and $i_{\rm O}$ represent the inclination 
of the inner and outer giant planet with respect to the plane of the binary, respectively. All other quantities are similar to 
those in Table 1.
\label{chartable}}
\tablehead{
\colhead{$i_{\rm I}$} & \colhead{$i_{\rm O}$} & \colhead{$N$} & \colhead{$m_{\rm p}$} & \colhead{$a_{\rm p}$} & \colhead{$\Delta {a_{\rm p}}$} & 
\colhead{$e_{\rm p}$} & \colhead{$i_{\rm p}$} & \colhead{$T_{\rm p}$} \\   
\colhead{(deg)} & \colhead{(deg)} & \colhead{} & \colhead{($m_E$)} & \colhead{(au)} & \colhead{(au)} & \colhead{} & 
\colhead{(deg)} & \colhead{(Myr)}  
} 
\startdata
0  & 0  & 1 & 1.31        & 0.59        & 0.01          & 0.07         & 2.72          & 7.40          \\
0  & 5  & 2 & 1.08 , 0.28 & 0.47 , 0.66 & -0.27 , -0.17 & 0.14 , 0.15  & 18.86 , 15.92 & 7.17 , 16.36  \\
0  & 10 & 1 & 1.18        & 0.46        & -0.06         & 0.25         & 17.3          & 18.75         \\
5  & 0  & 1 & 0.54        & 0.87        & -0.19         & 0.26         & 12.58         & 5.97          \\
5  & 5  & 0 &  \nodata    &  \nodata    & \nodata       & \nodata      & \nodata       & \nodata       \\
5  & 10 & 1 & 0.22        & 0.84        & -0.62         & 0.25         & 6.9           & 9.38          \\
10 & 0  & 0 &  \nodata    &  \nodata    &  \nodata      &  \nodata     &  \nodata      &  \nodata      \\
10 & 5  & 1 & 0.18        & 0.92        & -0.15         & 0.06         & 19            & 6.7           \\
10 & 10 & 0 &  \nodata    &  \nodata    & \nodata       & \nodata      & \nodata       & \nodata       \\
\enddata
\end{deluxetable*}

\begin{deluxetable*}{ccccccccc}
\tablecaption{Number, mass, and orbital properties of the final terrestrial planets in a 30 au, GG binary with an eccentricity of 0.2 
for different values of the orbital inclinations of the giant planets. Quantities $i_{\rm I}$ and $i_{\rm O}$ represent the inclination 
of the inner and outer giant planet with respect to the plane of the binary, respectively. All other quantities are similar to 
those in Table 1.
\label{chartable}}
\tablehead{
\colhead{$i_{\rm I}$} & \colhead{$i_{\rm O}$} & \colhead{$N$} & \colhead{$m_{\rm p}$} & \colhead{$a_{\rm p}$} & \colhead{$\Delta {a_{\rm p}}$} & 
\colhead{$e_{\rm p}$} & \colhead{$i_{\rm p}$} & \colhead{$T_{\rm p}$} \\   
\colhead{(deg)} & \colhead{(deg)} & \colhead{} & \colhead{($m_E$)} & \colhead{(au)} & \colhead{(au)} & \colhead{} & 
\colhead{(deg)} & \colhead{(Myr)}  
} 
\startdata
0  & 0  & 1  & 1.09        & 0.60        & -0.02          & 0.06          & 1.14      & 5.20        \\
0  & 5  & 1  & 0.22        & 0.57        & -0.26          & 0.59          & 8.84      & 8.6         \\
0  & 10 & 1  & 0.84        & 0.55        & -0.01          & 0.02          & 4.4       & 4.37        \\
5  & 0  & 1  & 0.40        & 0.56        & -0.19          & 0.05          & 17.74     & 10.4        \\
5  & 5  & 1  & 0.53        &  0.55       & -0.24          & 0.15          & 26.35     &   9.7       \\
5  & 10 & 1  & 0.16        & 0.87        & -0.04          & 0.02          & 8.84      & 1.65        \\
10 & 0  & 0  &  \nodata    &  \nodata    & \nodata        & \nodata       &  \nodata  &  \nodata    \\
10 & 5  & 0  &  \nodata    &  \nodata    & \nodata        & \nodata       & \nodata   & \nodata     \\
10 & 10 & 0  &  \nodata    &  \nodata    & \nodata        & \nodata       & \nodata   & \nodata     \\
\enddata
\end{deluxetable*}

\section{Summary and Concluding Remarks}

Continuing our study of the effects of secular resonances on the formation of terrestrial planets in moderately close binary stars, 
we have presented here the results of a large number of the simulations of the formation of these objects in a binary with two giant 
planets. Following Paper I, we considered a disk of planetesimals and planetary embryos, 
and simulated the late stage of terrestrial planet formation for different values of the mass, semimajor axis, and 
eccentricity of the secondary star, as well as the semimajor axes and inclinations of the giant planets. Results 
firmly confirm that as predicted by the general theory (Paper I), secular resonances are strongly suppressed by 
the perturbation of the secondary star and play no role in the formation of terrestrial planets.

\begin{figure*}[ht]
\center
\vskip 12pt
\hskip -14pt
\includegraphics[scale=0.35 , angle=-90]{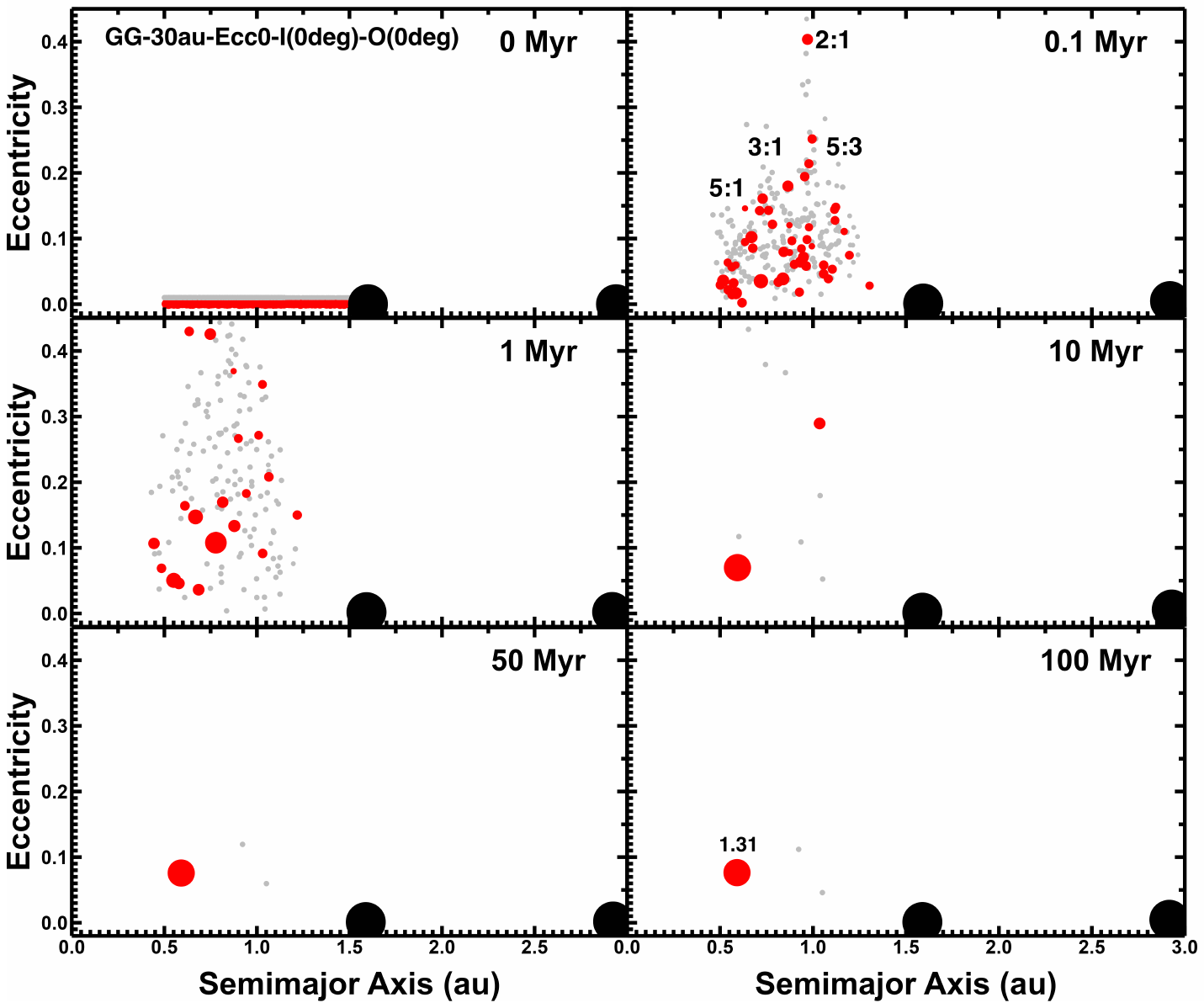}
\hskip -33pt
\includegraphics[scale=0.35 , angle=-90]{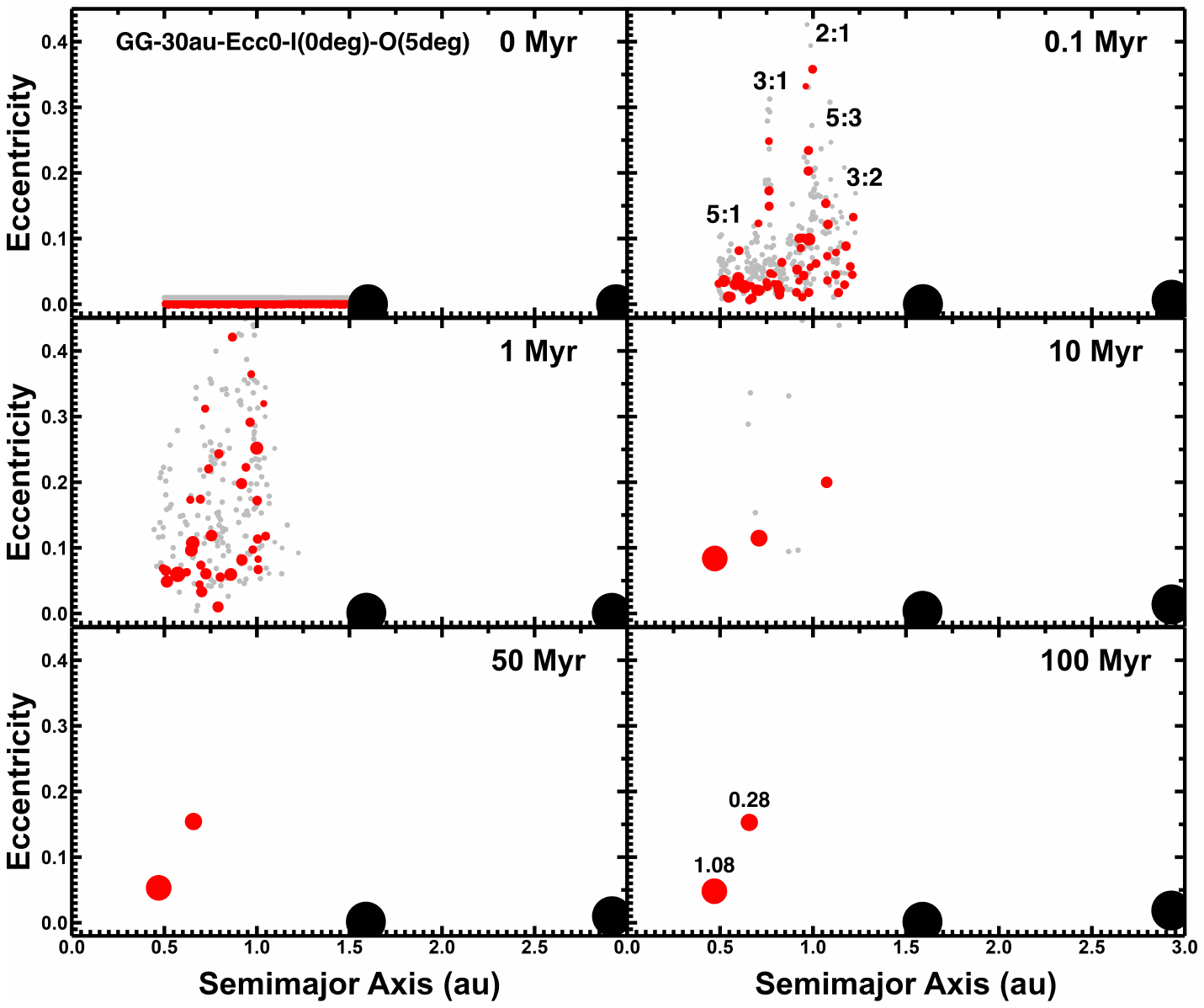}
\vskip -5pt
\hskip -14pt
\includegraphics[scale=0.35 , angle=-90]{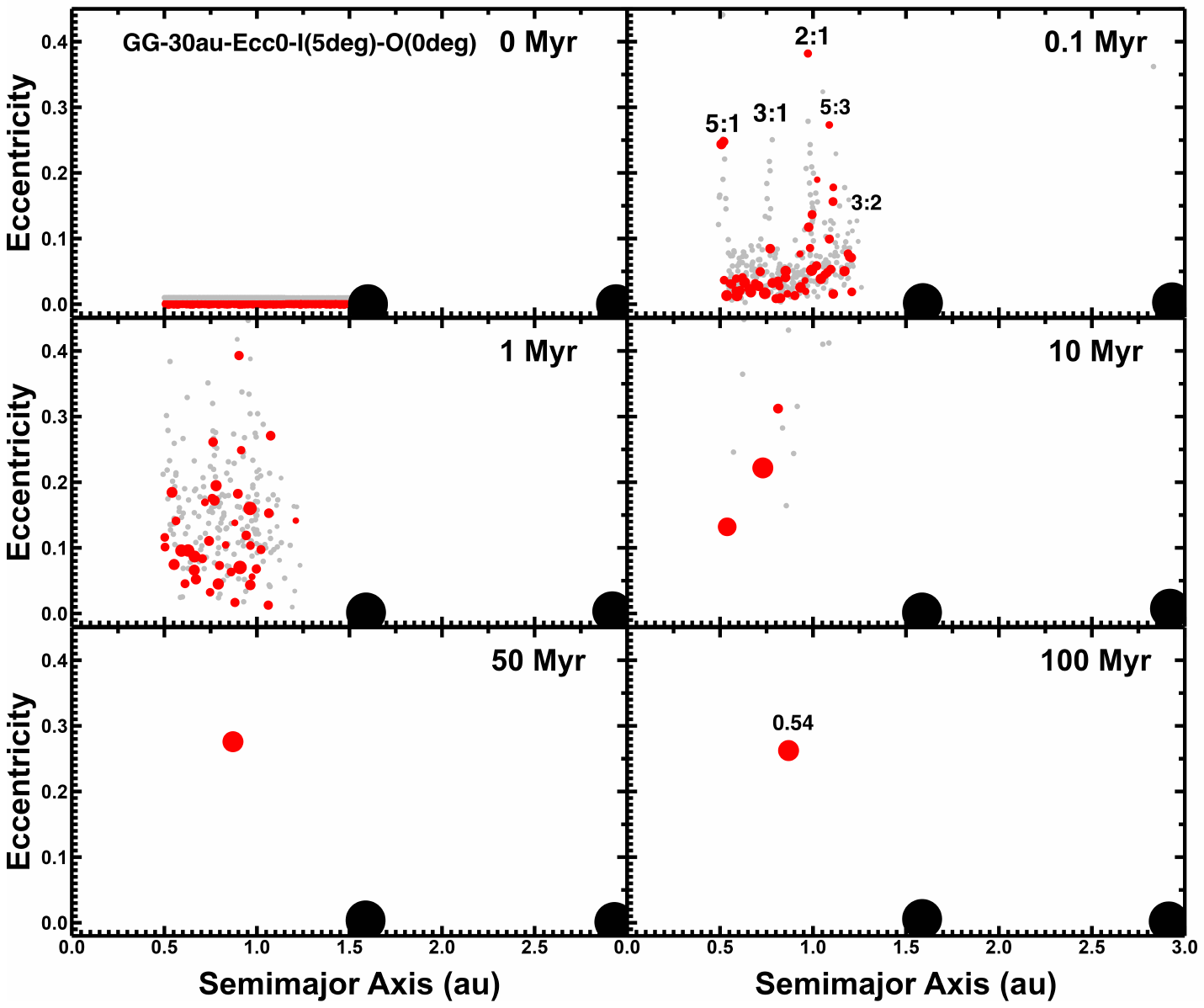}
\hskip -33pt
\includegraphics[scale=0.35 , angle=-90]{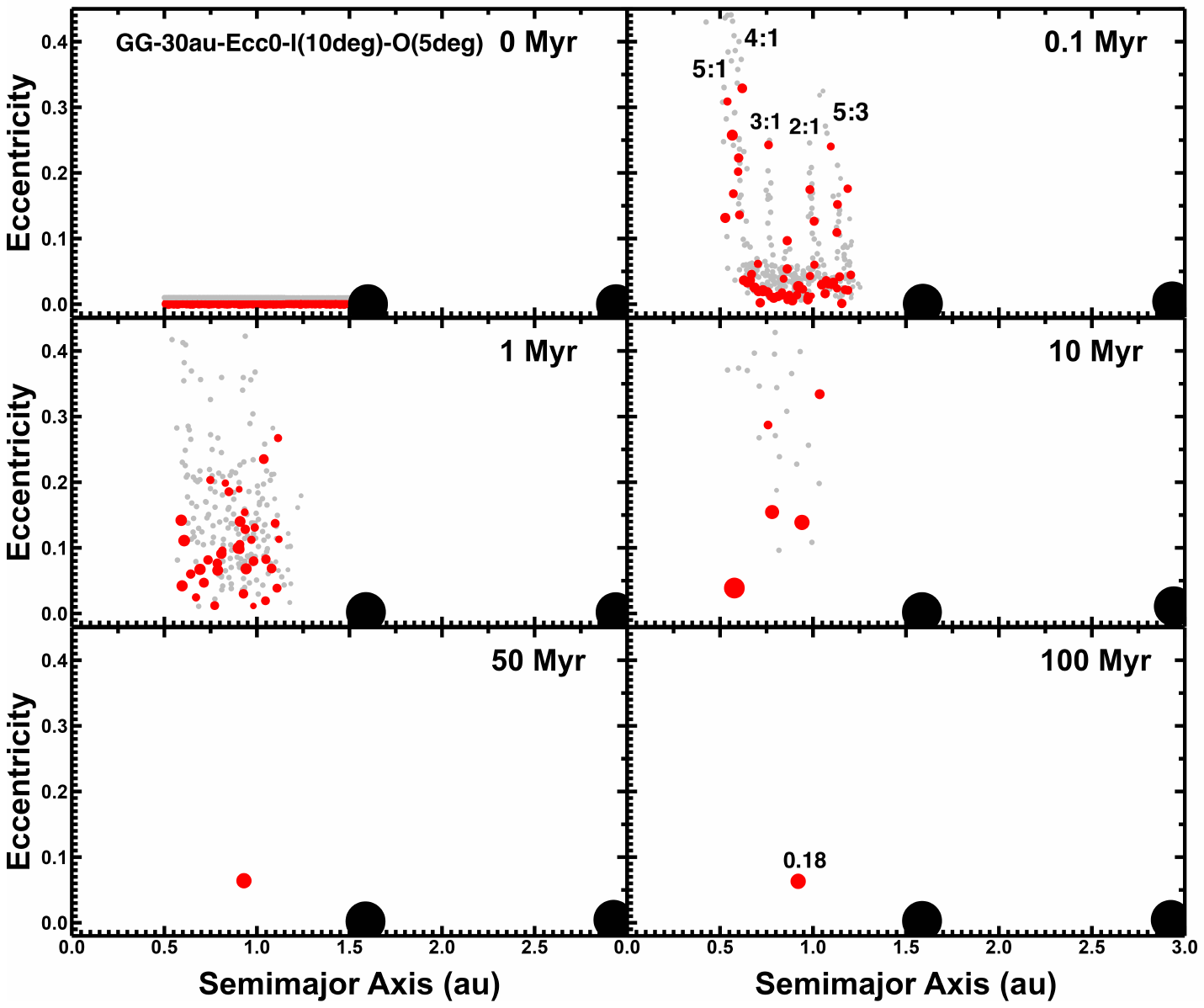}
\caption{Snapshots of the planet formation simulations in a circular, 30 au GG binary for different values of the orbital inclinations
of the two giant planets. The number above each red circle in the $t=100$ Myr panel indicates the final mass of that object 
in Earth mass. Note that in the naming of each simulation, I stands for the inner planet, O is the outer planet, and the numbers 
in parentheses correspond to their initial inclinations.}
\label{fig6}
\end{figure*}

Our simulations confirm the notion previously demonstrated by \citet{Haghighipour07} that when a planetary system is subject 
to the perturbation of additional stars, the giant planets act as a medium for transferring the perturbation of the stellar 
companions to the inner regions. Along that line, our integrations show 
that the dynamical evolution of the protoplanetary disk and the final outcome of the formation process are mainly and primarily driven by 
the mean-motion resonances of the inner giant planet. The increase in the orbital eccentricities of the disk's bodies 
due to the effect of these resonances, while removing some of the material from the disk, enhances the rate of the collisions 
among planetary embryos, furthering their growth to planetary bodies. This can be seen in Tables 1, 3, 5 and 6 where the majority of 
our simulations produced at least one terrestrial planet with a mass ranging between 0.6 and 1.7 Earth masses. 

When the combination of the perturbation of the secondary star and the effect of the mean-motion resonances allows the
system to form two planets, the outcome in most cases is an Earth-Mars analog. That is, the larger planet is Earth mass and 
the smaller planet has a mass in the range of the mass of Mars. This Earth-Mars correlation in our simulations resembles the Earth-Venus 
correlation in the simulations of the formation of terrestrial planets in our solar system, where Earth is an organic product of the 
formation process and Venus comes as a bonus after a strong orbital coupling is established 
between the growing seed of Venus and the growing 
Earth. In our simulations, however, there is no coupling between the growing Earth and Mars. The formation of Mars is purely random
and merely due to the occasional availability of more material in the outer part of the disk.

When simulations are carried out for different values of the semimajor axes of the giant planets, the results are
qualitatively similar but two additional features appear: the number of double-planet systems and their masses increase by increasing 
the semimajor axes
of the giant planets. The reason lies in the fact that in systems where the giant planets are placed in larger orbits, 
the strong, low-order mean-motion resonances also appear at larger distances and scatter material from the outer region of the disk, 
leaving more material in the
inner parts where the weaker, high-order resonances promote the collisional growth of protoplanetary bodies by increasing their 
orbital eccentricities and enhancing the rate of their collisions. Simulations demonstrate that Earth-mass and slightly more massive 
planets form routinely in systems where the outer giant planet is in an orbit larger than 3 au.

Results indicate that increasing the orbital inclinations of the giant planets negatively affects the formation outcome.
The analysis of the dynamical evolution of noncoplanar systems shows that the decoupling of eccentricity and inclination 
in these systems \citep{Li14} 
is not strong enough to counter the suppressing effect of the secondary star, and the secular resonances are still damped. 
However, increasing the inclinations of the two planets, especially that of the inner planet, strongly enhances the effect of the 
inner planet's mean-motion resonances to the extent that these resonances appear earlier and last longer than in the coplanar
simulations. These longer and stronger effects of the mean-motion resonances reduce the mass of the protoplanetary disk 
drastically, resulting in the decrease in the mass and number of final planets. Even at a low inclination of $10^\circ$, 
a nonnegligible number of the systems do not form any planets, or the mass of their final planet did not exceed a few 
time that of Mars. 

Our study demonstrates that in binary and multiple star systems, especially those with multiple giant planets, the additional 
stellar perturbation enhances the rate of terrestrial planet growth. This can be seen in all our simulations (see the last column
of Tables 1, 3, 5 and 6), where in almost all cases, the final planets formed within the first 20 Myr, and the large majority of 
them within the first 10 Myr. This short time of formation, which is in stark contrast to the time for the formation 
of Earth in the traditional simulations of solar system formation ($\sim 100$ Myr), while partially due to the proximity of the 
giant planets to the protoplanetary disk, is in large part the result of the perturbation of the secondary star. The latter can also 
be seen in the work of \citet{Haghighipour07} where the authors considered a Jupiter-mass planet at 5 au from a Sun-like primary star: 
the mere addition of a second star to their simulations decreased the time for the formation of Earth mass planets to 40 Myr.

In closing, our simulations indicate that, if planetesimal growth and the formation of planetary embryos proceed constructively around
a star of a moderately close binary with two (or, by the same token, multiple) giant planets, the formation of Earth analogs seems to 
be a natural product of the evolution of the disk. In other words, given the right conditions, terrestrial-class planets, including Earth 
and Mars analogs, can indeed form in moderately close binary systems with multiple giant planets. However, as predicted by the general theory 
(Paper I), and confirmed by our results, secular resonances do not play a significant role in the formation of these bodies.

We would like to thank the anonymous referee for critically reading our manuscript and for their useful suggestions and
recommendations. N.H. would like to acknowledge support through NASA grants 80NSSC18K0519, 80NSSC21K1050, and 80NSSC23K0270, 
and NSF grant AST-2109285.
We are deeply thankful to the Information Technology division of the Institute for Astronomy at the
University of Hawaii-Manoa for maintaining computational resources that were used for carrying out the numerical
simulations of this study.

\end{document}